\documentclass[12pt,preprint]{aastex}

\def\f24k{$F_{\rm MIPS}/F_{\rm photosphere}$}
\def\fk70{$F_{\rm MIPS}/F_{\rm photosphere}$}
\def\f24k{$F_{\rm MIPS24}/F_{\rm *,24\micron}$}
\def\fk70{$F_{\rm MIPS70}/F_{\rm *,70\micron}$}
\def\f24k{$F_{\rm MIPS}/F_{\rm *}$-24\micron}
\def\fk70{$F_{\rm MIPS}/F_{\rm *}$-70\micron}

\def\spit{{\it Spitzer}}

\def\70um{70~\micron}
\def\160um{160~\micron}
\def\24um{24~\micron}
\def\um{\micron}
\def\ld{L_{\rm dust}/L_{\star}}

\def\Mstar{M_{\star}}

\def\Tstar{T_{\star}}
\def\Lstar{L_{\star}}
\def\MEarth{M_\oplus}
\def\MSun{M_\odot}

\def\LSun{L_\odot}

\def\gapp{\lower 3pt\hbox{${\buildrel > \over \sim}$}\ }
\def\lapp{\lower 3pt\hbox{${\buildrel < \over \sim}$}\ }
\def\proptosim{\lower 3pt\hbox{${\buildrel \propto \over \sim}$}\ }

\def\arcsec{$^{\prime\prime}$}

\begin{document}

\title{New Debris Disks Around Nearby Main Sequence Stars: 
Impact on The Direct Detection of Planets}

\author{
C. A. Beichman$^{1}$, G. Bryden$^{2}$, 
K. R. Stapelfeldt$^{2}$, 
T. N. Gautier$^{2}$, 
K. Grogan$^{2}$, M. Shao$^{2}$, T. Velusamy$^{2}$, 
S.~M.~Lawler$^{1}$, M. Blaylock$^{3}$, G.~H. Rieke$^{3}$, J. I. Lunine$^{3}$, 
D. A. Fischer$^{4}$, G. W. Marcy$^{5}$, 
J.~S.~Greaves$^{6}$, 
M.~C. Wyatt$^{7}$, 
W. S. Holland$^{8}$, \& W. R. F. Dent$^{8}$ 
}

\affil{1) Michelson Science Center, California Institute of Technology, 
 Pasadena, CA 91125} 
\affil{2) Jet Propulsion Lab, 4800 Oak Grove Dr, Pasadena, CA 91109} 
\affil{3) Steward Observatory, University of Arizona, 933 North Cherry
 Ave, Tucson, AZ 85721}
\affil{4) Department of Physics and Astronomy, San Francisco State University, San Francisco, CA 94132}
\affil{5) Department of Astronomy, University of California, Berkeley,
 CA 94720}
\affil{6) School of Physics and Astronomy, University of St.~Andrews,
 North Haugh, St.~Andrews KY16 9SS, UK}
\affil{7) Institute of Astronomy, University of
 Cambridge, Cambridge, CB3 0HA, UK}
\affil{8) UK Astronomy Technology Centre, Royal Observatory,
 Edinburgh EH9 3HJ, UK}

\shorttitle{Debris Disks and Planet Detection}
\shortauthors{Beichman et al.}

\begin{abstract}

Using the MIPS instrument on the \spit\ telescope, we have searched
for infrared excesses around a sample of 82 stars, mostly F, G, and K
main-sequence field stars, along with a small number of nearby M
stars. These stars were selected for their suitability for future
observations by a variety of planet-finding techniques. These
observations provide information on the asteroidal and cometary
material orbiting these stars - data that can be correlated with any
planets that may eventually be found. We have found significant excess
\70um emission toward 12 stars. Combined with an earlier study,
we find an overall \70um excess detection rate of $13 \pm 3$\% for
mature cool stars.  Unlike the trend for planets to be found preferentially toward stars with high metallicity, the incidence of debris disks is uncorrelated with metallicity. By newly identifying 4 of these
stars as having weak \24um excesses (fluxes $\sim$10\% above the
stellar photosphere), we confirm a trend found in earlier studies
wherein a weak \24um excess is associated with a strong \70um
excess. Interestingly, we find no evidence for debris disks around 23
stars cooler than K1, a result that is bolstered by a lack of excess
around any of the 38 K1-M6 stars in 2 companion surveys. One
motivation for this study is the fact that strong zodiacal emission
can make it hard or impossible to detect planets directly with future
observatories like the {\it Terrestrial Planet Finder (TPF)}. The
observations reported here exclude a few stars with very high levels
of emission, $>$1,000 times the emission of our zodiacal cloud, from
direct planet searches. For the remainder of the sample, we set
relatively high limits on dust emission from asteroid belt
counterparts.

\end{abstract}


\section{Introduction}

A planetary system is characterized by the properties of its parent star, by the number and nature of its gas-giant and rocky planets, by the extent of its Kuiper and asteroid belts, and by the populations of gas and dust orbiting the central star. In the coming decade, astronomers will use a variety of techniques to address all these aspects of neighboring solar systems. Initial results for gas-giant planets are based on ground-based radial velocity searches. Eventually, nearby stars will be the targets of indirect and ultimately direct searches for terrestrial planets with the Space Interferometer Mission (SIM) and the Terrestrial Planet Finder (TPF). The \spit\ telescope is uniquely positioned to characterize the evolution, amount, structure and composition of the dust associated with Kuiper and asteroid belts around many types of stars, including those with and without the gas giant planets now being detected by the radial velocity technique. Guaranteed Time Observer (GTO) studies such as the FGK sample \citep{Beichman05mips, Bryden06a} and the Nearby Stars program \citep{Gautier06}, plus the FEPS Legacy project \citep{meyer04, kim05}, have conducted photometric surveys of about 200 nearby stars at 24 and \70um. The photometric survey discussed here uses \spit\ images at 24 and \70um to look for debris disks around an additional 82 stars, rounding out existing surveys of the closest stars.

As the Spitzer programs are completed, we will be able to carry out statistical investigations of the debris disk phenomenon in terms of the age, metallicity, and spectral type of parent stars. In particular, by nearly doubling the size of the existing sample of stars (relative to the on-going GTO/Legacy programs) we can hope to identify and improve the statistics of types of excess that appear to be rare based on existing IRAS or ISO observations, e.g. hot dust, extremely large disk to star luminosity ratios ($\ld$) around mature stars \citep{Fajardo00, Habing01, Spangler01}, or low mass stars. The incidence of excesses at the IRAS/ISO sensitivity level is about 15\% \citep{backman93, Bryden06a} so that a total \spit\ sample of 250-300 stars can hope to identify over 50 stars with excesses suitable for future study. Stars with hot excesses (peak wavelength $<$ \24um) are considerably rarer, 2-3\% \citep{Fajardo00, laureijs02, Beichman06irs}, so that a survey of a large number of stars is needed to generate a statistically meaningful sample. 

As the statistics of planets build up, we will be able to correlate the properties of debris disks (total mass, physical configuration, composition) with properties of number, location, and mass of planets. Much lower dust masses can be detected with \spit\ than was previously possible, particularly for solar-type and cooler stars. \citet{Beichman05mips} and \citet{Bryden06a} have shown that with \spit\ instruments, we can reach just a few times the fractional luminosity predicted for our own Kuiper Belt \citep[$0.3-5 \times 10^{-6}$;][]{backman93, stern96}. Determining how many mature stars like the Sun have Kuiper Belts comparable to our own is an important ingredient in understanding the formation and evolution of solar systems like our own \citep{levison03b}. 

Finally the \spit\ data will help us to understand the potential influence of zodiacal emission on the eventual direct detectability of planets. As highlighted in a number of TPF studies, including the {\it Precursor Science Roadmap for TPF} \citep{lawson04}, the level of exo-zodiacal emission can affect the ability of TPF to detect planets directly, particularly for extreme cases with much greater dust contamination than in the Solar System. A complete census of potential TPF stars will assist in the eventual selection of TPF targets by determining or setting a limit to the amount of exo-zodiacal emission around each star.

This paper focuses on the results of the \24um and \70um survey using
the Multiband Imaging Photometer (MIPS) instrument on \spit\
\citep[MIPS;][]{Rieke04}. After describing our target selection
(\S\ref{sample}), we present these MIPS observations in
\S\ref{observations}. As will be discussed below, a number of sources
in this sample appear to be extended. These are highlighted in
\S\ref{observations}, but are discussed in detail in a separate paper
\citep{Bryden06b}. Follow-up observations of sources with excesses
using the IRS spectrometer are just now being completed; these will
also be detailed in a later paper. In \S\ref{ldsec} we discuss how
our MIPS flux measurements constrain the dust properties in each
system. For the systems identified here as having IR excess,
combined with those from \citet{Bryden06a}, \S\ref{parameters}
attempts to find correlations between the dust emission and system
parameters such as stellar metallicity, spectral type, and
age. Finally, in $\S$\ref{imp} we assess the influence of debris disks
on the detectability of planets. 

\section{Stellar Sample}\label{sample}

Our sample is based on work carried out by radial velocity search
teams \citep[e.g.][]{Marcy04} and by the SIM and TPF Science Teams to
identify the most suitable targets for the indirect or direct
detection of terrestrial-mass planets (1-10 $\MEarth$). One target
list consists of the 100 stars in the SIM Tier-1 sample which will be
the most intensively observed stars in the two SIM projects dedicated
to finding planets around nearby stars \citep{Marcy02sim, Shao02}. \footnote{The merged, high priority target list for these projects is available at \hfil\break http://astron.berkeley.edu/$\sim$gmarcy/sim\_draft.html.} 
Since the absolute astrometric signal from a planet scales as 3~$\mu {\rm sec} \ (D_{\star}/{\rm pc})^{-1} \times (a_{\rm planet} /{\rm AU}) \times (M_{\rm planet}/\MEarth) \times (\MSun/\Mstar)$, the SIM teams are concentrating on the some of the closest, lower mass stars for their deepest surveys for terrestrial planets. Thus, the SIM list includes a number of late K and M stars not included in the other \spit\ samples or in TPF lists.

We also draw from a number of lists prepared by Science Working Groups
for the TPF-Coronagraph (TPF-C) and TPF-Interferometer (TPF-I) missions. Although the TPF lists are not definitive given the indeterminate status of the project, the outline of the sample is clear \citep{BeichmanPPV, TraubIAU}. We start with F0 - M5 stars of luminosity classes IV or V and refine the list by making a few simple assumptions about the nature of planetary systems and the properties of TPF. Specifically, we 1) exclude stars with binary companions within 100 AU as being inimical to the formation or stable evolution of planetary systems; 2) require that the angular extent of the habitable zone \citep{kasting1993}; $\sim$1 AU for a 1 $\LSun$ luminosity star, scaled by the square root of the stellar luminosity) exceed 50 milliarcsec; and 3) impose an outer distance cutoff of 25 pc (although we allowed a few F0-F5 stars at distances as great as 30 pc to bring up their numbers). To enable good measurements with \spit\ we rejected stars with high levels of stellar and/or cirrus confusion based on examination of IRAS maps. 

Comparison of potential SIM and TPF targets in cirrus-free sky with the \spit\ Reserved Object Catalog (as of November 2003) showed 81 stars with spectral types ranging from F0 to M3.5, as listed in Table~\ref{basictable}. One more star, GL 436, was added through a Director's Discretionary Time proposal after the discovery of a planet in this system was announced \citep{butler04}. Binary companions within the 82 fields of view have also been included as secondary targets; six such companions are identified as bright enough for clear detection in both the 24 and \70um images.\footnote{The \24um image of HD~265866 has what appears to be an equal-brightness binary companion located 40\arcsec\ NW of the target primary star. However, there is no visible or near-IR neighboring source. In fact, the second \24um source is a chance alignment with a passing asteroid. Software specifically developed for locating asteroids relative to the \spit\ observatory \citep[part of the Horizons package;][]{Giorgini05} identifies this object as asteroid \#11847 \citep[``Winckelmann'': H=13.4, a=2.67 AU, e=0.065, i=10.23;][]{Bowell05}.} The divergent proper motion of HD~48682B, an M0 star 30\arcsec\ to the NE of HD~48682, rules out a physical association between the two stars. Thus, HD~48682B is not included in this sample. The true binarity of the other six neighboring sources is verified via their Hipparchos distances and space motion measurements. Angular separations in these systems range from 10\arcsec\ to 100\arcsec, with projected orbital separations between 100 and 1000 AU. Data for the companions are listed separately at the end of Tables~\ref{basictable} and \ref{mipstable}. With their inclusion, our total sample contains 88 stars within 82 targeted fields. 

\begin{figure}
\begin{center}
\includegraphics[width=4.7in,angle=-90]{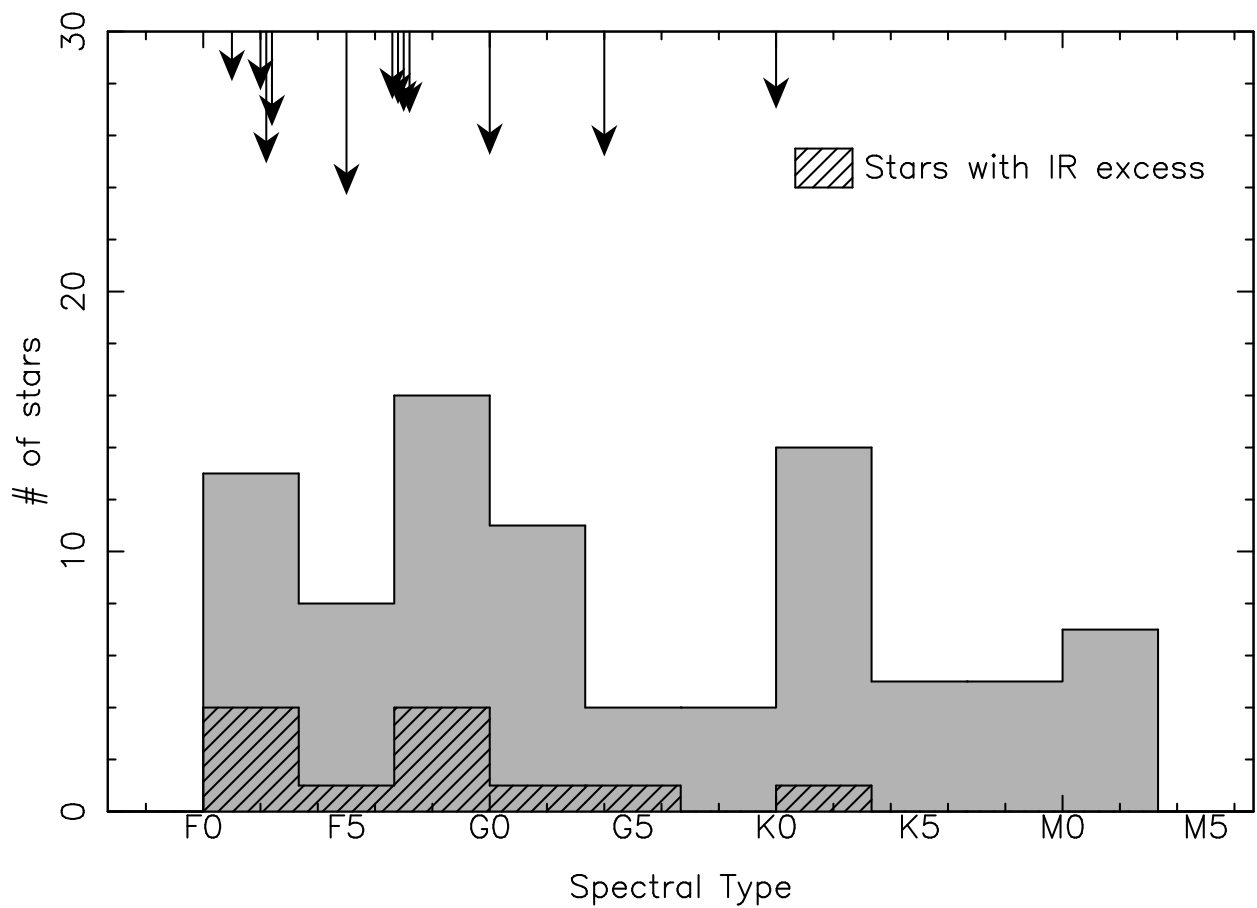} 
\end{center}
\caption{Spectral type distribution for stars in this SIM/TPF sample.
The spectral types of stars found to have \70um excess are 
highlighted within the histogram (slant-hash) and are individually
flagged with arrows at the top
of the plot. The length of each arrow is an
indicator of the strength of \70um excess relative to the stellar
photosphere. We find that \70um excess is more readily detected around
early type stars.} 
\label{sptype}
\end{figure}

\begin{figure}
\begin{center}
\includegraphics[width=4.7in,angle=-90]{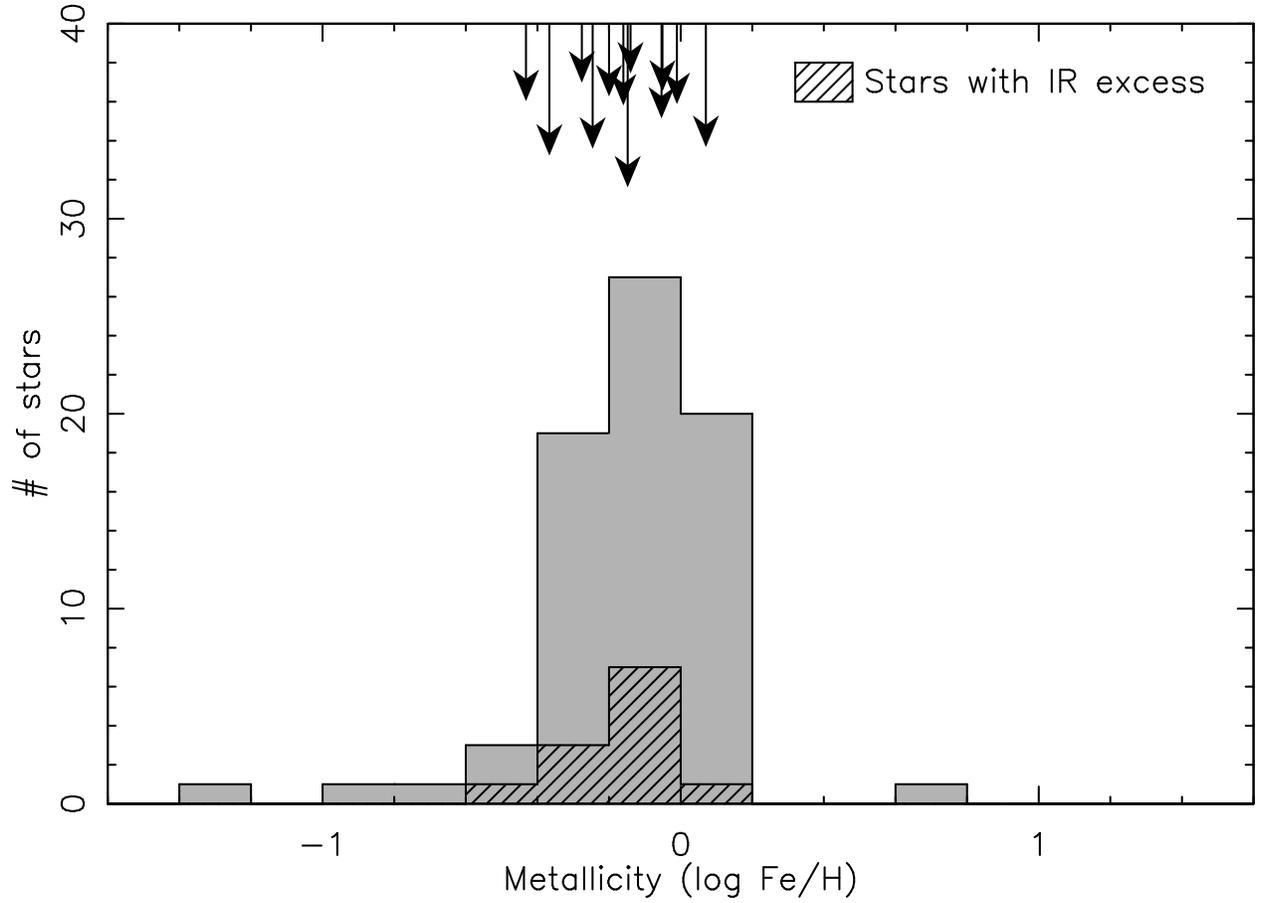} 
\end{center}
\caption{Metallicity distribution for stars in this SIM/TPF sample.
The ages of stars with \70um excess are 
highlighted within the histogram (slant-hash) and are individually
flagged with arrows at the top
of the plot. The length of each arrow is an indicator of the strength
of \70um excess relative to the stellar photosphere. There is no
correlation between metallicity and the detection of IR excess.}
\label{metals}
\end{figure}

Binned by spectral type, the SIM/TPF sample consists of 37 F stars, 19
G stars, 24 K stars, and 8 M stars. Typical distances range from 10 to
20 pc, closer for M and K stars and farther for earlier spectral types.
Figure~\ref{sptype} shows the overall distribution of observed
spectral types. Some basic parameters of the sample stars are listed in
Table~\ref{basictable}, most importantly age and metallicity, which
are also shown as histograms in Figures~\ref{ages} and \ref{metals}.
There is no explicit target selection based on stellar age or
metallicity, but known planet-bearing stars have been specifically
included in a couple cases. Of this sample, only two stars (GJ 436 and
HD~147513) are already known to have planets; most of the other stars with
planets are either too faint, lie in cirrus-contaminated regions, or
are already observed in other \spit\ programs \citep[e.g.][]{Beichman05mips}. 

In this paper we first discuss the 88 primary and secondary stars and then add in the stars observed in \citet{Bryden06a} to increase the size of the sample for some of statistical discussions. 

\section{\spit\ Observations}\label{observations}

All stars were observed with MIPS at \24um and, with one exception
(HD~265866), at \70um. In order to help pin down their stellar
photospheres, four M stars - GJ 908, HD~36395, HD~191849, and
HD~265866 - were also observed with the IRAC camera in subarray mode
at all four of its wavelengths (3.6, 4.5, 5.8, and 8.0 \micron). Seven
stars identified as having IR excess were observed with IRS, the \spit\ 
spectrograph, as follow-up observations, as detailed in a future paper. 

\subsection{Data Reduction}

\subsubsection{MIPS Observations}

Overall, our data analysis is similar to that previously described in
\citet{Beichman05mips} and \citet{Bryden06a}. At \24um, images were created from the raw data using the DAT software developed by the MIPS instrument team \citep{Gordon05}. At \70um, images were processed beyond the standard DAT software to correct for time-dependent transients, corrections
which can significantly improve the sensitivity of the measurements
\citep{Gordon06}. For both wavelengths, aperture photometry was performed using
apertures sizes, background annuli, aperture corrections, and
instrument calibration as in \citet{Beichman05mips}. We find that the
target locations in the \24um images are consistent with the telescope
pointing accuracy of $<$1\arcsec\ \citep{werner04}. As such, we use
the \24um centroid as the target coordinates for both wavelengths.
Special consideration is made for the 6 resolved binaries in our sample.
Instead of our standard method of aperture photometry with a
surrounding sky annulus, the emission at the two stars' locations is fit 
with the instrument's point spread function (PSF). We find that for binary stars with small angular separations, simultaneously fitting of their overlapping PSFs results in much improved photometric accuracy. The
agreement between PSF fitting and aperture photometry (with
appropriate aperture correction) for isolated stars is excellent \citep{Gordon05}. For all of the stars, the MIPS flux and 
noise measurements are listed in Table~\ref{mipstable}. 

\subsubsection{IRAC Observations}

The IRAC sub-array images of the four M stars were reduced following the technique described by the FEPS Legacy team \citep{Carpenter06}. 
The pixel sizes are corrected for distortion and a pixel-phase
correction is made to channel 1. Stellar fluxes are measured within an aperture of 10 pixels (=12$^{\prime\prime}$), with a background annulus from 10 to 20
pixels. The photometric measurements each star at the four IRAC
wavelengths are listed in Table~\ref{irac}. 

\subsection{Photospheric Extrapolations and Limits on \24um Excess}

To determine whether any of our target stars have an IR excess, we compare the measured photometry against predicted photospheric levels. A detailed description of our stellar atmosphere fitting, as applied to F5-K5 stars, is presented in the appendix of \citet{Bryden06a}. The stars observed here, however, span a greater range of spectral types than previously considered.
In particular, our sample contains late K and M type stars with numerous broad molecular features for which the stellar models \citep{Kurucz03} begin to lose their accuracy.

\begin{figure}
\begin{center}
\includegraphics[width=4.7in,angle=-90]{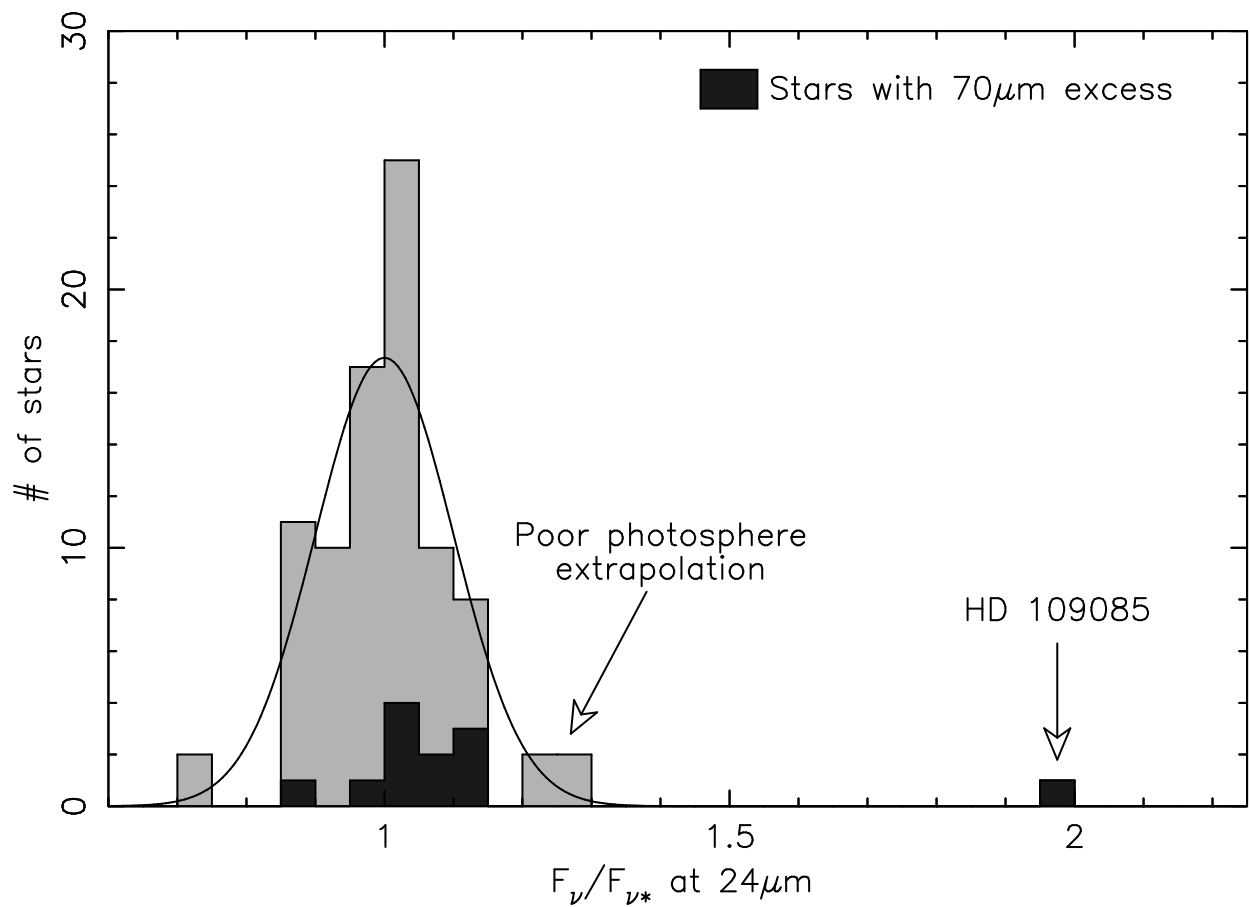} 
\end{center}
\caption{Distribution of \24um fluxes relative to the expected
photospheric values. A Gaussian distribution with 10\% dispersion
({\it solid curve}) is shown for comparison. One star (HD~109085, a
star previously identified as having excess emission) clearly stands
out from the main population. 
The broad dispersion within this population is due to a variety of
factors. Some stars have poor estimates of the stellar flux at \24um
due to poor near-IR data or photospheric models, particularly for the set
of late K and M stars marked 'poor extrapolation' in the figure.
The spread of values is also increased by sources 
with true, weak excesses (at the level of $\sim$10\% above the stellar
photosphere). 
Stars with excesses at a longer wavelength (\70um) are shown with
black shading.} 
\label{f24k}
\end{figure}

This accuracy can be directly assessed by examining how well the
observed flux levels match those predicted. We use the ratio, \f24k,
to assess the photospheric extrapolation using the fact previously
established in \citet{Bryden06a}, \citet{Beichman06irs}, and earlier
references cited therein, that excesses at \24um are rare ($\sim$1\%).
Figure~\ref{f24k} shows the distribution of this ratio. 
After excluding one outlying star with a strong \24um and
\70um excess (HD~109085), the 88 flux measurements at \24um have an
average \f24k of 1.01. The dispersion of \f24k in Figure~\ref{f24k} is
0.10, which is relatively large compared to the previous result for just F5-K5
stars \citep[0.07;][]{Bryden06a}. We identify three causes for this
larger dispersion: 

\subsubsection{Quality of Near-IR Photometry}

The SIM/TPF sample contains a number of nearby stars that are brighter than the stars in the FGK survey. Stars brighter than about K$_s=4$ mag have saturated 2MASS measurements resulting in large photometric uncertainties ($\sim$0.25 mag). For several of these stars, particularly the early F type stars which are the brightest in the sample, Johnson $K$-band photometry is available in the literature with much better accuracy ($\sim$0.05 mag) than the saturated 2MASS values, but worse than the best 2MASS values ($\sim$0.03 mag). In such cases, the saturated 2MASS values are supplanted by the better data. However, the uncertainty in the near-IR photometry for the remaining 2MASS-saturated stars causes difficulty in extrapolating to longer wavelengths and results in a greater dispersion than when only stars with high quality 2MASS data are used, $\sigma$(\f24k)= 0.09 vs. 0.07 for types F5-K5. 

\subsubsection{Intrinsic Variability}

Stellar variability between the epochs of the Spitzer data and the photometry used to estimate the photospheric contribution could account for some of the dispersion in \f24k. To investigate this possibility we examined the Hipparcos photometry for the 66 stars of our sample for which these data are available (ESA 1997). Only three stars (GL 436, HD 79211 and HD 265866) showed a scatter in $mag(Hp)$ in excess of 0.02 mag while the vast majority had scatter less than 0.01 mag. In particular, none of the ten stars younger than 1 Gyr and thus possibly more variable than the rest of the sample, showed variability above this level. GL 436 and HD 265866 are faint V$\sim$10 mag M stars so that the level of Hipparcos scatter is not significant. The large scatter for HD 79211 (0.23 mag) is due to multiplicity and is also not significant. As discussed below, one non-Hipparcos star, HD~38392, shows a low level of variability and a correspondingly larger deviation in \f24k.

\subsubsection{Quality of Photospheric Models}

The ability to extrapolate from visible and near-IR photometry to MIPS
wavelengths appears to be an issue for spectral types later than the
F5-K5 range used in the FGK survey. For all stars with accurate 2MASS
data, Figure~\ref{kv24} plots the directly observable $K_s -$[24]
color, a quantity independent of the stellar atmosphere models. With
the exception of the M stars, all of the averages are consistent with a
constant color of $K_s -$[24] $\simeq 0.02 \pm 0.02$. Most interestingly, an apparently abrupt transition occurs between the late K stars and M stars, with the average $K_s - $[24] color jumping up $\sim$0.4 magnitudes for the cooler stars. This trend of redder $K_s - $[24] for later spectral types was first noticed by \citet{Gautier06} whose M star data are shown for comparison. 

We next consider the ratio of the observed flux at \24um to that
predicted by photospheric models (\f24k) as a function of spectral type.
Solar-like stars (types F5-K4) have an overall average of
\f24k $=0.98\pm0.01$ with a dispersion of 5\% among the stars 
with good 2MASS data and excluding stars with excess emission at \70um
(identified in \S\ref{70umsection}). For F0-F4 stars, the observed fluxes are
marginally higher than those predicted, with an average \f24k $= 1.03
\pm 0.02$. For late K stars with good 2MASS observations the observed
fluxes are consistently below expectation with an average \f24k of
$0.87 \pm 0.03$. Since the observed K-[24] color is flat (Fig.~\ref{kv24}), this offset is likely a fault of the photospheric modeling or of our fitting procedure. Not surprisingly, the models have the greatest difficulty with the M stars which have average \f24k = $1.16 \pm 0.06$. This difference between measured and predicted fluxes for the M stars remains even if NextGen \citep{Hauschildt99} models are used instead of Kurucz models. However, with an accurate determination of each star's effective temperature and with more advanced stellar models \citep[PHOENIX;][]{brott05}, \citet{Gautier06} were able to fit the \24um colors of M stars. 

Thus, knowing that these trends in \f24k exist and may ultimately be explained with better modeling, we can compare each star with the average K$_s$-[24] color within its spectral type bin (Fig.~\ref{kv24}) to look for dust excesses. With this methodology, we find no evidence for a \24um excess toward any of our M stars. This negative result is strengthened when IRAC photometry is available. For the 4 M stars with IRAC data (Table~\ref{irac}), the inclusion of 3.5-8.0 \micron\ fluxes into the fit modifies the average of \f24k from 1.32 with a dispersion of 0.21 to \f24k=0.92 with a dispersion 0.04, confirming that the M stars' \24um fluxes are consistent with emission from the stellar photosphere alone. 

\subsubsection{Presence of a Weak Excess at 24 $\micron$}

Finally, the third reason for increased dispersion in the \f24k values,
in addition to poor near-IR photometry and less accurate stellar
photospheres or model fitting for late type stars, is the presence of
weak but real excess emission from dust toward some stars. In \S\ref{70umsection} we will identify some of our target stars as having strong excess emission at \70um. Only one of these objects, HD~109085\footnote{An
excess was first detected around HD~109085 (= $\eta$ Crv) by IRAS
\citep{Aumann88,Stencel91} and subsequently with SCUBA at sub-mm wavelengths
\citep{Sheret04,Wyatt05}. Consistency between the IRAS flux at 25
\micron\ and the MIPS \24um flux measured here depends strongly on the
application of color corrections which are functions of the assumed
dust temperature. For dust temperatures around 200-400 K and assuming
the photospheric value given in Table~\ref{mipstable}, we find good
consistency between the two measurements.} (the labeled value in
Figure~\ref{f24k}) also has an immediately obvious IR excess at \24um. 

Taken in composite, however, the stars with \70um excesses tend 
to have a weak \24um excess. Considering only F0-K5 stars with good near-IR photometry, those with \70um excess have an average \f24k value 9\% higher than 
those without. A similar general trend was previously noticed by
\citet{Bryden06a} and was confirmed in the IRS spectra of individual
objects with \70um excess, which tend to rise above the stellar
photosphere longward of 25 \micron\ \citep{Beichman06irs}. Combining
the F0-K5 stars in this sample with those from \citet{Bryden06a}, we
are no longer limited by small number statistics and the correlation
between \70um and \24um excess becomes significant at the 3-$\sigma$
level. The average \24um excess for stars with \70um excess is
0.079$\pm$0.026 times the stellar flux. 

To assess the significance of a possible \24um excess on a star-by-star basis, we define the parameter $\chi_{24}$ which corresponds to the $n-\sigma$ significance of any deviation from the expected photospheric value: 
\begin{equation}\label{chi24eq}
\chi_{24} \equiv \frac{F_{24} - F_{\star}}{\sigma_{24}}
\end{equation}
where $F_{24}$ is the measured flux, $F_{\star}$ is the expected
stellar flux, and $\sigma_{24}$ is the noise level, all at \24um.
A similar definition follows for \70um (equation~[\ref{chi70eq}]).
We take the noise to be the larger of either 4\% for sources with good 2MASS or Johnson data or 8\% for sources with poor near-IR photometry. 
These values are based on the dispersions in \f24k for the stars without \70um excesses. Ignoring the previously discussed late K and M stars with poor photospheric extrapolations, we find that the deviations from photospheric emission skew sharply to positive values for stars with \70um excess
({\it black shading} in Fig.~\ref{f24k}). Using this analysis we identify statistically significant 24 $\micron$ excesses accompanying a stronger 70 $\micron$ excess around two stars: HD~25998 (3.6 $\sigma$) and HD~40136 (3.2
$\sigma$) in addition to HD~109085 discussed earlier. At slightly
lower significance we find hints of a 24 $\micron$ excess for HD~199260
(2.7 $\sigma$) and HD~219482 (1.9 $\sigma$) which the accompanying 70
$\micron$ excess suggests could be real. 

A number of other stars show strong deviations from photospheric
values without an accompanying \70um excess: the deviant \f24k
values of the M stars have already been discussed and attributed
to poor photospheric extrapolation; HD~38392 has an apparent 30\%
excess at \24um which we attribute to the difficulty of obtaining
an accurate measurements due to a) saturated 2MASS measurements, b)
proximity to a nearby, bright companion (HD 38393) and c) the possible
variability of the star itself at the 5\% peak-to-peak level
\citep{Nitschelm00}. Finally, HD~23249 and HD~55892 have $2<\chi_{24}<3$ and $\chi_{70}<2$. These deviations could simply be statistical fluctuations or they could be hints of an excess like that seen toward HD 69830 which is prominent only in the 8-34 $\micron$ region but not at \70um \citep{Beichman05irs}. Without additional data, e.g. IRS spectra, we cannot assess the reality of the
excesses around these last two stars. 

\subsection{Detection of \70um Excess}\label{70umsection}

\begin{figure}
\begin{center}
\includegraphics[width=4.7in,angle=-90]{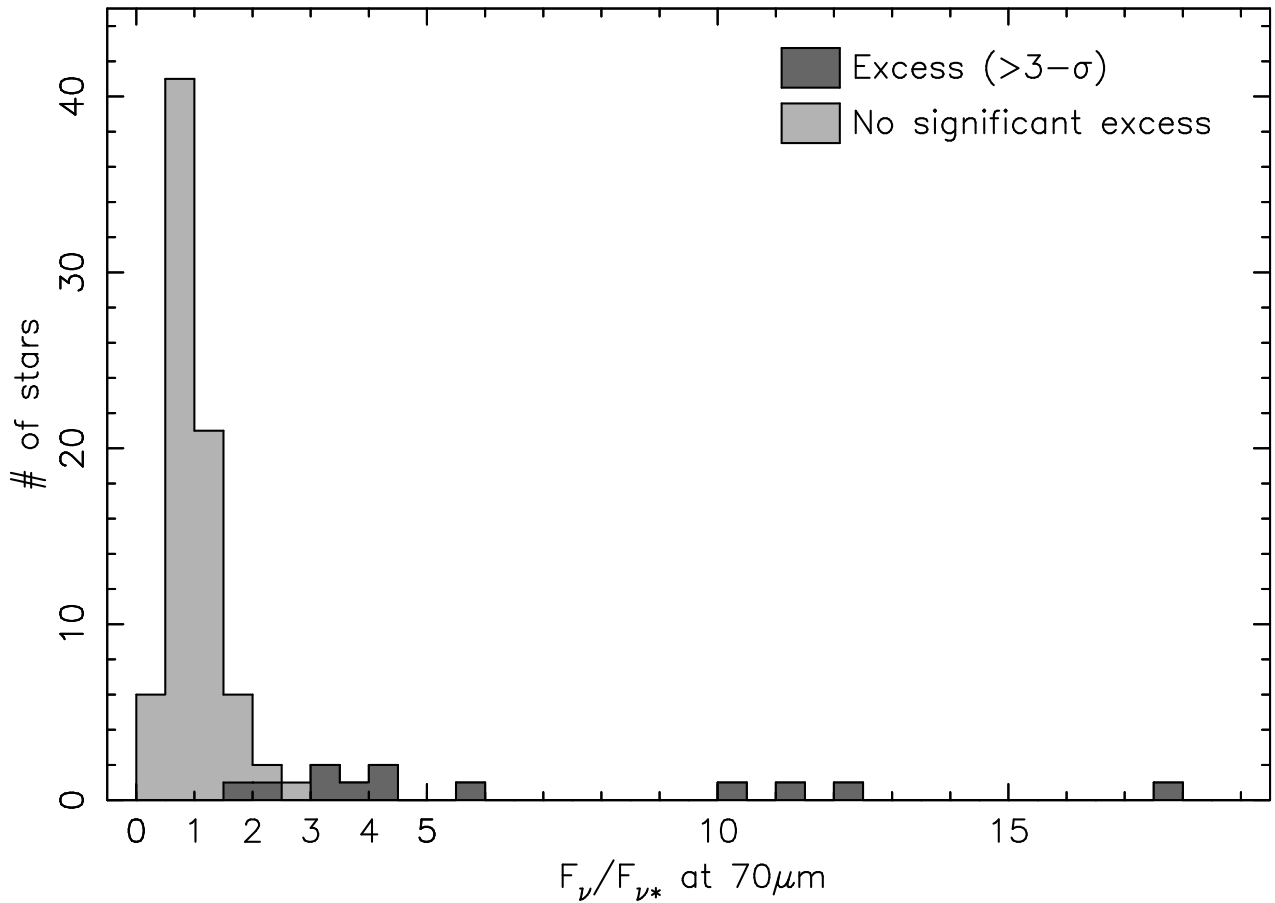} 
\end{center}
\caption{Distribution of \70um fluxes relative to the expected
photospheric values. While most stars cluster around unity
(consistent with emission from the star alone) many show a high degree
of excess emission attributable to circumstellar dust.} 
\label{f70k}
\end{figure}

Having used \24um fluxes to test the accuracy of our stellar
photosphere predictions, we next consider the frequency and strength
of excess emission at \70um. The distribution of \70um flux densities
relative to the expected photospheric values is shown in Figure~\ref{f70k}. Unlike the tight distribution of flux ratios at \24um, many stars have \70um flux densities much higher than expected from the stellar photosphere alone. In several cases, the flux is more than an order of magnitude greater than expectation. Twelve of these stars will be identified in the following as having
statistically significant IR excess. Excluding these stars with excesses and those with signal-to-noise ratio (S/N) $<3$, the average ratio of MIPS flux to predicted photosphere is $|$\fk70$| =1.02 \pm 0.05$, consistent with the overall calibration. 

The dispersion in the \70um data is $\sim$40\% (excluding the stars with
excesses), considerably higher than that in the \24um data. An
analysis of the noise levels in each individual field is required to
assess whether the IR excesses are statistically significant. Many
contributions to the overall error budget must be considered including
those arising from stellar photosphere modeling, instrument
calibration, sky background variation, and photon detector noise. At
\70um, the calibration uncertainty and the background noise within 
each image are considerably larger than at \24um. On top of an assumed
calibration uncertainty of 15\%, we directly measure the standard
deviation of the background flux when each field is convolved with
our chosen aperture size. This background noise, which ranges from
$\sim$2 to 20 mJy with a median of 3.7 mJy, is due primarily to
extragalactic source confusion and cirrus contamination, rather than
photon noise, and hence cannot be greatly reduced by additional integration time \citep[for a more detailed analysis of the \70um noise levels, see][]{Bryden06a}. Based on this measured background noise, we determine the S/N for each star, as listed in Table~\ref{mipstable}. Despite the high level of noise in some fields due to cirrus contamination and/or background galaxies, 72 out of the 87 stars in our sample with \70um data are detected with
signal-to-noise ratio greater than 3. The median S/N for all of our
target stars is 6.6, excluding the sources identified as having excess
emission (which have a median S/N of over 20). 

Adding both background noise and calibration error together gives us
a total noise estimate for each \70um target. In Table \ref{mipstable} we list these noise levels, along with the measured and the photospheric fluxes, for each observed star. We use these noise estimates to calculate $\chi_{70}$ which corresponds to the $n-\sigma$ significance of any deviation from the expected level of photospheric emission: 
\begin{equation}\label{chi70eq}
\chi_{70} \equiv \frac{F_{70} - F_{\star}}{\sigma_{70}}
\end{equation}
where $F_{70}$ is the measured flux, $F_{\star}$ is the expected
stellar flux, and $\sigma_{70}$ is the noise level, all at \70um.
Based on this criterion, we find that 12 out of 88 stars have a
3-$\sigma$ or greater excess at \70um: HD~25998, HD~38858, HD~40136, HD~48682, HD~90089, HD~105211, HD~109085, HD~139664, HD~158633, HD~199260, HD~219482, 
and HD~219623. In a sample of 88 stars there should be fewer than 1
star with a spurious excess on purely statistical grounds. Although
cirrus or extragalactic confusion could produce spurious excesses,
careful examination of each of the 70 \micron\ images suggests that

this is unlikely in the vast majority of cases. 
For example, the 70
\micron\ emission is well centered on the 24 \micron\
positions, typically within$\sim$1\arcsec. 
A number of weak excesses could have escaped detection
under these criteria. Observations at higher sensitivity or at higher
spatial resolution will be needed to identify these. 
The detection rate of \70um excess within this sample is $14 \pm 4$\%;
combined with the sample of \citet{Bryden06a}, this gives an
overall detection rate of $13 \pm 3$\% for cool stars.

Four of these stars have been previously identified as having excess
emission: HD~40136 \citep[=$\eta$ Lep;][]{Aumann88,Mannings98}, HD~48682 
\citep[=$\psi$ 5 Aur;][]{Aumann91,Sheret04}, HD~109085 \citep[=$\eta$
Crv;][]{Aumann88,Wyatt05}, and HD~139664 \citep[= g Lup;][]{Walker88,Habing96,Kalas06}. The eight newly discovered
IR-excess stars mostly have \70um fluxes less than 100 mJy, too dim to have been be detected by IRAS. The notable exception is HD~105211
which has a very strong \70um flux ($\sim$500 mJy), but lies
near a bright infrared source (CL Cru); with a separation of 2.4$^\prime$,
this source is easily resolved in the MIPS image (Figure~\ref{MIPSimage}),
but still contaminates the large IRAS beam. For the stars without any significant excess emission, 3-$\sigma$ upper limits on possible excess flux typically range from 0.2 to 1.0 times the stellar flux, with a median of upper limit of 0.6 $F_{\star}$. 

Although the telescope resolution at \70um is relatively poor (FWHM of
$\sim$17\arcsec), several of these sources appear to be slightly
extended in the MIPS images (marked with superscript {\it f} in Table
\ref{mipstable}). As discussed in a separate paper \citep{Bryden06b}, examination of the images of these marginally resolved sources do not indicate contamination by background objects, e.g. cirrus or galaxies, but rather 
that the objects possess to be truly extended disks. For one of the five Spitzer-resolved disks, HD~139664, a Hubble Space Telescope image shows the same orientation of the disk in the visible as in the infrared \citep{Kalas06}.
At distances of 10-20 pc, the resolved disks have apparent radii of
100's of AU. As discussed below, maintaining a temperature of $\sim$50 K at these distances (warm enough to emit strongly at \70um), requires relatively small grains with low emissivities. 

\section{Properties of the Detected Dust}\label{ldsec}

Beyond our initial goal of detecting IR excesses, we are interested in
determining the properties of the dust in each system - its
temperature, luminosity, mass, size distribution, composition, orbital
location, etc. For the 12 stars with significant IR excess, Table~\ref{excesstable} lists the excess \70um emission and measurements of or limits to the \24um excess. Seven of the 12 excess sources have an excess measured only at one wavelength (\70um), restricting our ability to place limits on key quantities. The observed flux can be translated into the total dust disk
luminosity relative to its parent star only when some assumption is
made for the dust temperature \citep[e.g.][]{Beichman05mips,Bryden06a}. The minimum disk luminosity as a function of \70um dust flux density is
obtained by setting the emission peak at \70um (or, equivalently, setting $T_{\rm dust} = 52.5$ K): 
\begin{equation}\label{ldeq}
\frac{L_{\rm dust}}{L_{\star}}({\rm minimum}) = 10^{-5} 
\; \left(\frac{5600 \; {\rm K}}{T_{\star}}\right)^3
\; \frac{F_\nu(70\, \mu m, {\rm dust})}{F_\nu(70\, \mu m, \star)}
\end{equation}
Based on this equation, a minimum $\ld$ is calculated for each of our
target stars identified as having an IR excess at only \70um (Table
\ref{excesstable}). For the other IR-excess stars, we also have at least a rough (2-$\sigma$) detection of the \24um excess, as discussed in the previous section. With two wavelengths of excess measurement, the dust
emission can be fit with a representative dust temperature, otherwise
only upper limits can be obtained (Table~\ref{excesstable}; also see
Figure~\ref{Tbytype} below). Figure~\ref{sed} shows a spectral energy distribution (SED) for two of these stars, HD 219482 and HD 40316, fit with temperatures of 170 and 80 K respectively. In the cases with a fit dust temperature, Table~\ref{excesstable} lists the ratio of the
integral under the stellar ($T_\star$) and dust ($T_{\rm dust}$)
blackbodies as a proper, rather than minimum, estimate of $\ld$: 
\begin{equation}\label{ldeq2470}
\frac{L_{\rm dust}}{L_{\star}} = \; \left(\frac{T_{\rm dust}}
{T_{\star}}\right)^4 \left(\frac{e^{x_{\rm dust}}-1} {e^{x_\star}-1}\right) 
\; \frac{F_\nu(70\, \mu m, {\rm dust})}{F_\nu(70\, \mu m, \star)}
\end{equation}
where $x \equiv {h\nu}/{kT}= {205.7 \, {\rm K}}/{ T}$ at \70um.

\begin{figure}
\begin{center}
\includegraphics[width=4.5in,angle=-90.]{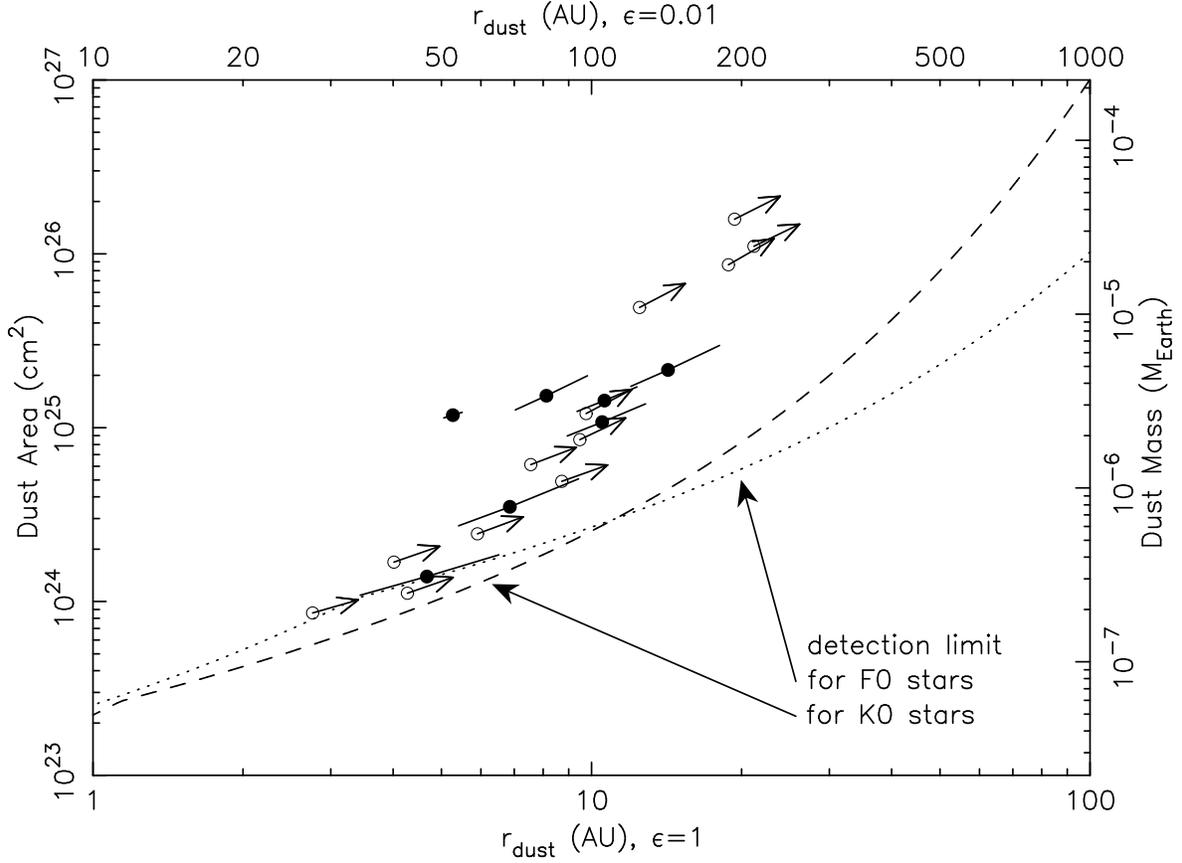} 
\end{center}
\caption{Area and dust mass estimates for stars with \70um excess
emission. In addition to our 12 stars with excess, the 7 excess stars
from \citet{Bryden06a} are also shown. Dust temperatures (Table~\ref{excesstable}) are translated into orbital radii assuming either large blackbody grains (bottom axis) or small grains with emissivity 0.01 (top axis). Stars with excess measured at both 24 and \70um are shown as solid
points, while those with only upper limits for the dust temperature
are shown as open circles. 
Error bars are added to each point based on the 1-$\sigma$
uncertainties in the dust temperature; 
for systems with upper limits on the dust temperature, 
an arrow is plotted with length/direction based on an assumed 10\%
uncertainty in the temperature.
Dust masses (right axis) are calculated assuming a typical grain size
of 10 \micron. Both dust area and mass are calculated under the
assumption of blackbody grains (unity emissivity); for an emissivity
of 0.01, both area and mass are a factor of 100 larger. The detection
limits, which depend on the stellar temperature, are shown for a F0
star (dotted line) and for a K0 star (dashed line).} 
\label{rdust3}
\end{figure}

For each of the stars with excess emission (plus those from \citet{Bryden06a},
Figure~\ref{rdust3} shows the total dust area and radial location of the emitting material. Despite the expectation that only large grains should be seen around mature systems due to loss mechanisms such as Poynting-Robertson drag and radiation pressure, small grains must be considered as a serious possibility given their presence in a mature star like HD~69830 \citep{Beichman06irs} and because of the large extent of at least some the dust disks. Thus, the orbital location of the emitting material can only be calculated if some assumption is made for the dust's emissivity; small grains are less efficient at
emitting infrared radiation, resulting in a higher temperature for a
given orbital radius. For a given dust temperature, the orbital radius decreases with emissivity as $\epsilon^{-0.5}$. As such, for each calculated dust temperature or upper limit in Table~\ref{excesstable}, two locations are plotted - the location if the emitting material is large blackbodies (lower axis) and the location if it is small grains with emissivity = 0.01 (upper axis).

The dust area and mass have a similar ambiguity based on the unknown
dust size/emission properties. The dust area in Figure~\ref{rdust3} (left axis) is calculated under the assumption of blackbody grains (unity emissivity); a lower emissivity would increase the dust area in direct proportion to $\epsilon$. Lastly, the dust mass (right axis) is based on an assumed typical
grain size of 10 \micron. An unconstrained amount of mass is 
contained in the larger parent bodies whose collisions produce the
emitting dust.

For stars with no detected emission, 3-$\sigma$ upper bounds on the
\70um fluxes lead to upper limits on $\ld$ as low as a few times
$10^{-6}$, assuming a dust temperature of $\sim$50 K
(Table~\ref{mipstable}). Although we cannot rule out cold dust at $\gapp$100 AU, we are placing constraints on dust at Kuiper Belt distances to $\sim$10-100 times the level of dust in our solar system. The constraint on asteroid
belt-type dust is less stringent, $\sim$100-1000 times our zodiacal emission.

\subsection{Comparison with Sub-mm Observations}

The dust properties can be further constrained with sub-mm flux measurements. When available, longer wavelength data can help place lower limits on dust temperature, upper limits on the dust luminosity, and, with some assumption on the grain emissivity, outer limits on the disk extent. For most of our stars with IR excess, large amounts of cold dust emitting at longer wavelengths cannot be ruled out, but three of the \70um excess stars have been observed at
450 and 850 \micron\ with JCMT/SCUBA, with two detections (HD~48682
and HD~109085) and one upper limit \citep[HD~139664;][]{Sheret04}.
Combining their sub-mm data with the infrared fluxes from IRAS,
\citet{Sheret04} modeled the SEDs for these stars, obtaining dust
temperatures of 99 and 85~K for HD~48682 and HD~109085, respectively.

\subsection{Comparison with Visible Observations}

A nearly edge-on disk around HD~139664 has recently been detected at
visible wavelengths using the Hubble Space Telescope (HST)
\citep{Kalas06}. The HST image of the disk shows a well defined inner
edge around 83 AU and an outer edge that extends out to about 109
AU. If the dust associated with the \70um emission is located in this
ring, then a very large surface area of 10 $\micron$ grains would be
required for the IR emission, since dust at this distance would have
an equilibrium temperature of 30-35~K using a standard radial power
law for grain temperature \citep{Beichman06irs}. Smaller, 0.25
$\micron$ grains, on the other hand, give temperatures of 69 and 77~K
at the ring boundaries (cf.\ our 3-sigma upper limit of 78 K;
Table~\ref{excesstable}). Using a simple relationship for dust mass,
$M_{dust}={4 \over3} \rho a_{grain} { {D2 F_\nu(dust)} \over {Q_{abs}
B_\nu(T_{dust})}}$, and standard silicate absorption efficiencies,
$Q_{abs}$ \citep{Draine84, Beichman06irs}, yields a mass of
$2.4\times10^{-3} M_\oplus$ in large grains or $1.6\times10^{-4}
M_\oplus$ in small grains, where we have taken a grain density of
$\rho$=3.3 gm cm$^{-3}$ and a distance, $D=17.5$ pc, for this star. 

The radiative blowout size for grains around an F5 star is
approximately 1 \micron\ 
(\S V.B.1 in \citet{backman93};
\citet{burns79}). In contrast to the spherical distribution of small
grains seen toward Vega \citep{Su05} which is probably due to a
recent catastrophic event, small grains should be quickly ejected from
the presumably  stable ring system of HD~139664. IRS spectroscopy
and/or millimeter spectroscopy would help distinguish between the
large and small grain models.

\section{Correlation of Excess with System Parameters}\label{parameters}

To understand the origin and evolution of infrared excess, we now
consider the properties of the sample stars and how they correlate
with excess detection. Specifically, we examine the correlation with three variables: metallicity, age, and spectral type. These parameters are listed for each star in Table~\ref{basictable}. Where appropriate we merge the present sample with that of \citet{Bryden06a} to improve the significance of any statistical conclusions.

\subsection{Metallicity}

Table~\ref{basictable} lists the metallicity information obtained from
the literature for each of our target stars (number of independent [Fe/H]
estimates, their average, and their r.m.s.\
scatter). Figure~\ref{metals} shows a histogram of these metallicity
values, the majority of which range between -~0.4 to +0.1 dex with a
mean value of somewhat below solar. The stars with IR excess are
identified with vertical arrows with the length of each arrow 
proportional to the strength of \70um excess. We find no correlation
between metallicity and IR excess. The average [Fe/H] is
-0.15$\pm$0.03 for all the observed stars and -0.17$\pm$0.04 for the
stars with excess - an insignificant difference.  

\begin{figure}
\begin{center}
\includegraphics[width=4.7in,angle=-90]{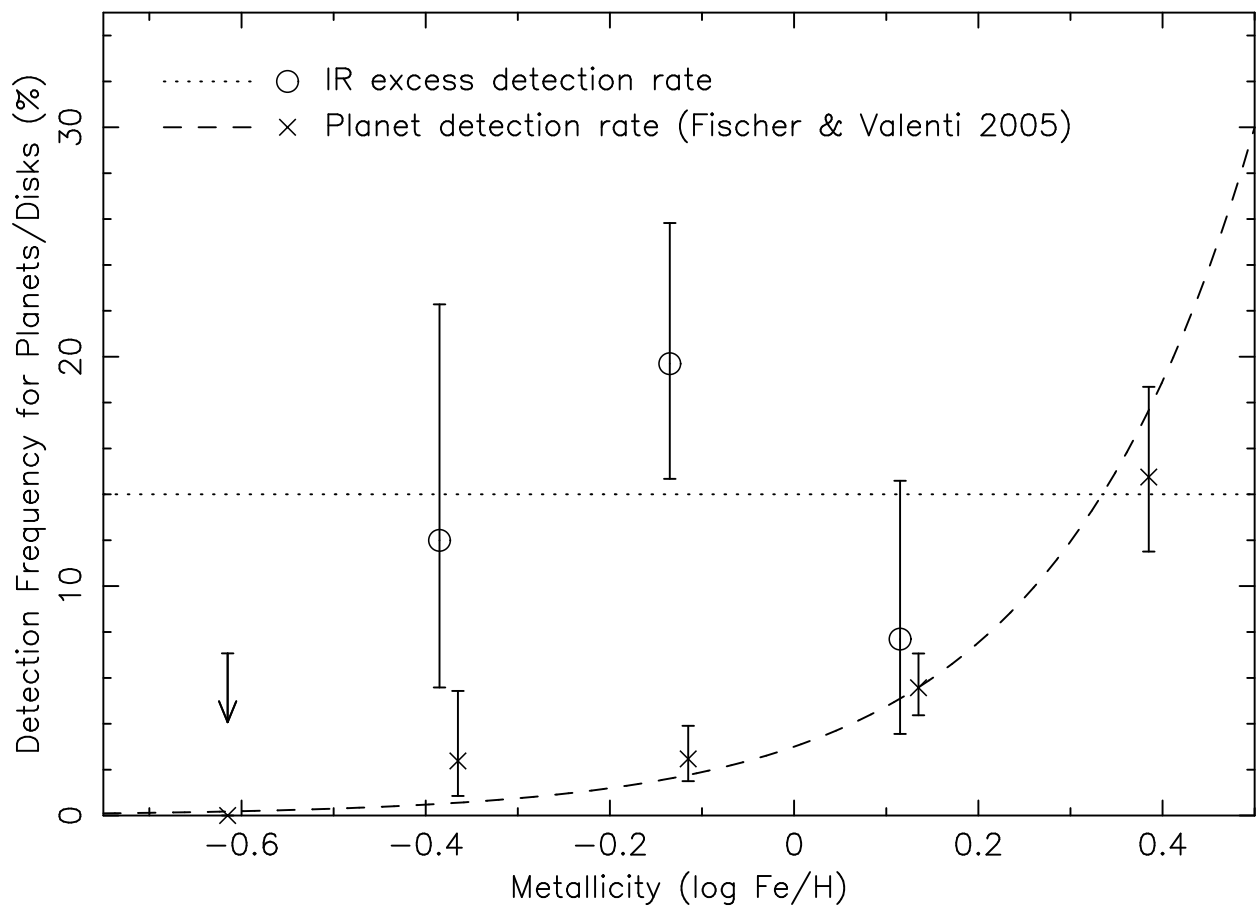} 
\end{center}
\caption{Detection frequency of planets and IR
excess as a function of stellar metallicity.
For the stars presented in this paper, plus those of \cite{Bryden06a},
the detection rate of \70um excess emission is shown as open circles.
The distribution has no trend in metallicity;
all points are consistent with the average detection rate for the entire sample
(14\%; dotted line).
This is in contrast to the dependence of the planet detection rate
on metallicity for a similar sample of nearby stars (x marks), 
which \citet{fischer05} fit with a
metallicity-squared relationship (dotted line).}
\label{fischcompare}
\end{figure}

These data are combined with \citet{Bryden06a} sample to show the
fractional incidence of disks as a function of [Fe/H] (Figure
\ref{fischcompare}). A $\chi^2$ test shows that the distribution of
disks in our three metallicity bins with significant number of stars
(-0.75 $<$[Fe/H]$<$ 0.0) is indistinguishable from flat. The lack of
correlation between IR excess and metallicity is in sharp contrast
with the well known correlation between extrasolar gas giant planets
and host star metallicity \citep{gonzales97,santos01}. In particular,
\citet{fischer05} find that the probability of harboring a
radial-velocity detected planet increases as the square of the
metallicity (Figure \ref{fischcompare}). Although one might suspect
giant planets and debris disks to be related, a similarly strong
correlation between dust and metallicity can be confidently ruled
out. A $\chi^2$ comparison between the disk and planet distributions
in same 3 [Fe/H] bins suggest that there is only a 0.3\% probability
of these being drawn from the same distribution.  

\begin{figure}
\begin{center}
\includegraphics[width=4.7in,angle=-90]{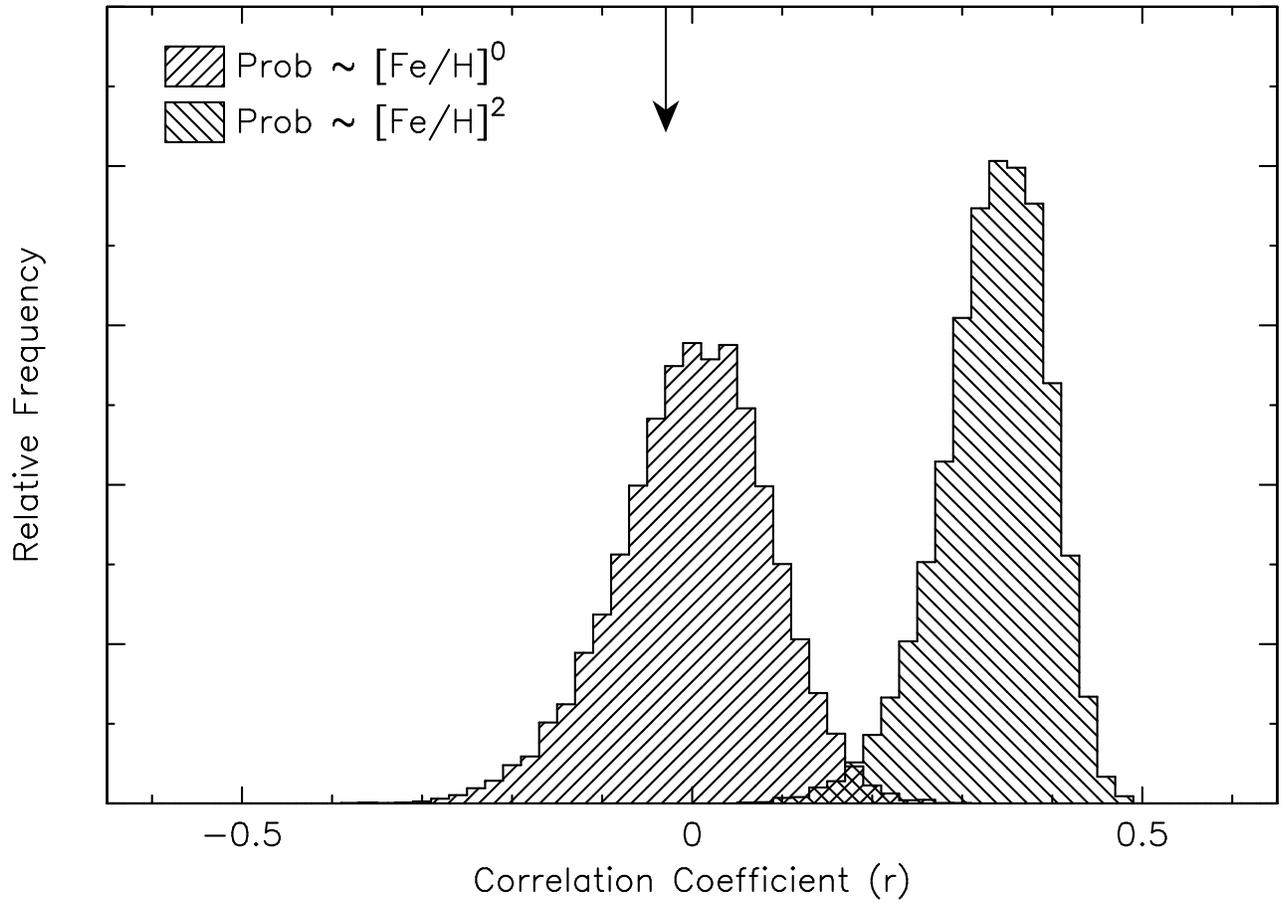} 
\end{center}
\caption{Distribution of $r$ correlation coefficients for two series
of monte carlo simulations.
Stars were selected either completely randomly (left histogram) or
proportionate to their metallicity squared (right histogram), 
the relationship observed for planet-bearing stars.
The arrow at the top shows the value of $r$ observed within our data, 
strongly inconsistent with the planet-metallicity relationship.
}\label{finalfig}
\end{figure}

The lack of correlation is further confirmed via Monte Carlo
simulations.  Again using our dataset combined with that of 
\citet{Bryden06a} (giving a total of 19 excess stars out of 151 observed),
the correlation coefficient, $r$, is calculated for 10,000 random samples of stars. The histogram of the resultant $r$ values is shown in
Figure~\ref{finalfig} under two different assumptions.
In one case, the stars with excess are chosen randomly (left histogram
centered on $r$=0); in the other, the stars are chosen with weighting
proportional to their metallicity squared (right histogram with
average $r$=0.33). The correlation coefficient observed within our data (vertical arrow) is inconsistent with the strong metallicity dependence observed within planet-bearing stars.

This lack of correlation may reflect the different formation histories 
of giant planets and debris disks. The accretion of gas onto a giant
planet requires a large solid core to form first, favoring a higher
metallicity disk, whereas dust emission indicates the presence of
smaller planetesimals that might be able to form in all disk environments.
Another explanation may be that debris disks in high metallicity
systems initially contain more material, but that over time all disks
grind down toward similar masses \citep[e.g.][]{Dominik03}. 
In this case, the detection of strong IR emission is a reflection of a
recent stochastic collision, rather than the disk's initial conditions (see \citet{Bryden06a} for further discussion).

\subsection{Age}

Collisions in a debris disk continually grind down the larger planetesimals, while the smallest dust can be removed by Poynting-Robertson drag and radiation pressure. One would assume that the overall disk mass must decline with time and, as expected, a correlation between stellar age and IR excess is observed, with debris disks more commonly identified around younger stars. While studies concentrating on stars younger than 1 Gyr find a strong trend \citep{Spangler01, Rieke05}, among nearby solar-type field stars the correlation is relatively weak \citep{Bryden06a}. In both cases, the evolution of the dust does not appear to be a steady decline. Observations of A stars find an overall decline in the
average amount of \24um excess emission on a $\sim$150 Myr time scale, but the large variations on top of this trend suggest that sporadic collisional events are able to dramatically increase the amount of dust even at late stages in the disk's evolution \citep{Rieke05}. As a result of these collisions, even old stars can have strong IR emission \citep{Habing01,decin00,Bryden06a}. 

\begin{figure}
\begin{center}
\includegraphics[width=4.7in,angle=-90]{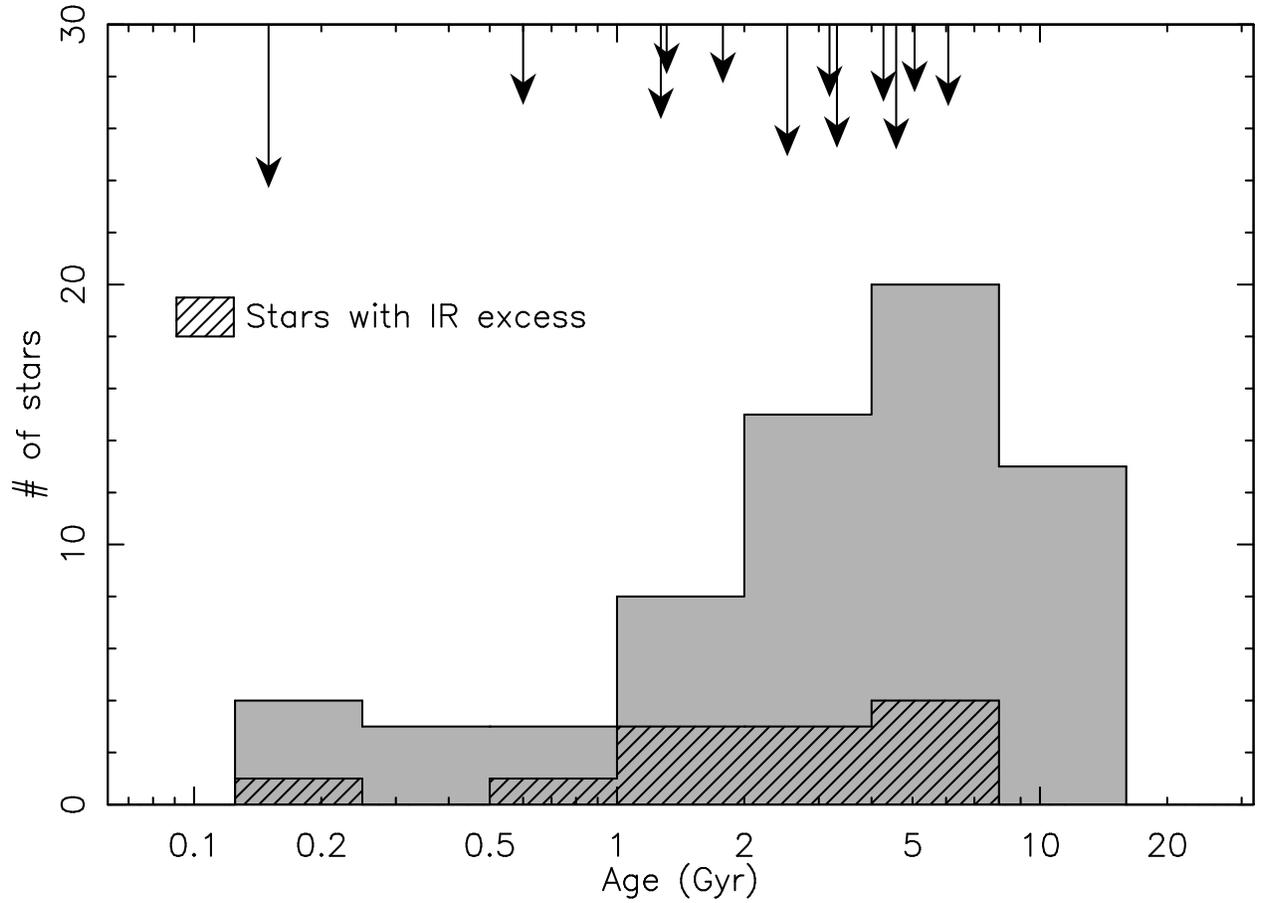} 
\end{center}
\caption{Age distribution for stars in this SIM/TPF sample.
The ages of stars with \70um excess are 
highlighted within the histogram (slant-hash) and are individually
flagged with arrows at the top
of the plot. The length of each arrow is an indicator of the strength of \70um excess. There is a weak correlation between the detection of IR excess and the stellar age, with no stars older than 7 Gyr having excess emission.}
\label{ages}
\end{figure}

Figure~\ref{ages} shows the resultant histogram of stellar ages. The ages for our main sequence stars are difficult to determine, with uncertainties in many cases of at least a factor of two. Where possible, we use ages based on Ca II H\&K line emission from the large compilation of \citet{Wright04}. Otherwise an average of values found in the literature is used. If the star was inferred to be young due to kinematic properties \citep{Montes01}, we adopted that
age. Table~\ref{basictable} lists the age data for each star. Although
our target selection criteria do not explicitly discriminate based on
stellar age, young stars (ages less than 1 Gyr) are not well
represented in our sample due to their infrequent occurrence within
$\sim$25 pc of the Sun. 

As in our earlier survey of nearby main-sequence stars \citep{Bryden06a}, 
the stars with excess in this survey (marked with arrows in
Figure~\ref{ages}) have a weak but noticeable correlation with stellar
age. No stars older than 7 Gyr have a significant amount of excess
emission. The average age of stars with IR excess is 4.0$\pm$0.6 Gyr,
compared to 5.6$\pm$0.4 Gyr for the sample as a whole. As discussed in the next section, these trends are present in the combination of this sample with the \citet{Bryden06a} data.

\subsection{Spectral Type}\label{sptypsec}

Observations of the general characteristics of debris disk as a
function of spectral type are potentially a powerful tool for
understanding the physical mechanisms responsible for the evolution of
debris disks. The disk properties should be directly related to the stellar mass and luminosity in several ways. The mass of the protostellar disk from
which the debris formed, for example, probably depends on the 
parent star's mass, as does its dynamical time scale. The stellar
luminosity, though, is undoubtedly more important for debris disk
characteristics, exerting a strong influence on the typical particle
size ($r_{\rm blowout} \propto \Lstar$) as well as its temperature
($T_{\rm dust} \propto \Lstar^{0.2-0.25}$). There are also observational
biases linked to the brightness of the star, with cool dust seemingly
easier to distinguish around hotter stars. 

The minimum $\ld$ based on the \70um flux, for example, is strongly
dependent of stellar temperature (in eq.~[\ref{ldeq}], detectable
$\ld$ is proportional to $\Tstar^{-3}$). Hotter stars emit a lower
fraction of their luminosity at infrared wavelengths, allowing for
better contrast at those wavelengths. But while equation~(\ref{ldeq})
is an observationally well-defined quantity, it contains no knowledge
of the underlying disk physics. Naively, it appears to dictate a
strong relationship between detectability and spectral type, i.e.\ it
is easier to detect dust around hotter stars, but this may be
misleading. For lack of any other information, the equation assumes
that the dust emission peaks at \70um, thereby measuring the minimum
$\ld$. This assumed SED shape corresponds to a fixed dust temperature
of $\sim$50 K for all disks. One can instead consider disk models with
a more physically motivated dependence on spectral type. Instead of 
assuming a constant dust temperature, \citet{Habing01}, for example,
assume the same dust {\it location} for all disks; in their models,
the dust resides at 50 AU independent of spectral type. In this case,
dust temperature decreases with $\Tstar$. In contrast with a simple
reading of equation~(\ref{ldeq}), the Habing et al.\ models have $\ld$
more or less directly proportional to $F_{{\rm dust}}/F_{\star}$ for
stars of type G and earlier. As in equation~(\ref{ldeq}), lower
stellar temperature makes it more 
difficult to detect dust emission relative to the stellar photosphere,
but in Habing et al.'s models this difficulty is offset by a lower dust temperature for cooler stars, increasing the dust's \70um emission. 

In Figure~\ref{sptype}, the spectral types with IR excess stars are
flagged with vertical arrows. A clear trend is readily apparent, with
excess more frequently detected around earlier type stars. 
The detection rate drops from nearly 30\% for the earliest type stars
down to 0\% for M stars. In fact, no stars with spectral type later than K0 are found to have excess emission (a sample of 23 stars without excess). 
This is consistent with previous survey results that considered only part of the spectral range covered here. Our survey of F5-K5 stars \citep{Bryden06a} found a detection rate of 13\% within this limited spectral range, while a sample of
$\sim$30 images of nearby M stars yielded none with IR excess at \70um \citep{Gautier06}. 

\begin{figure}
\begin{center}
\includegraphics[width=4.7in,angle=-90]{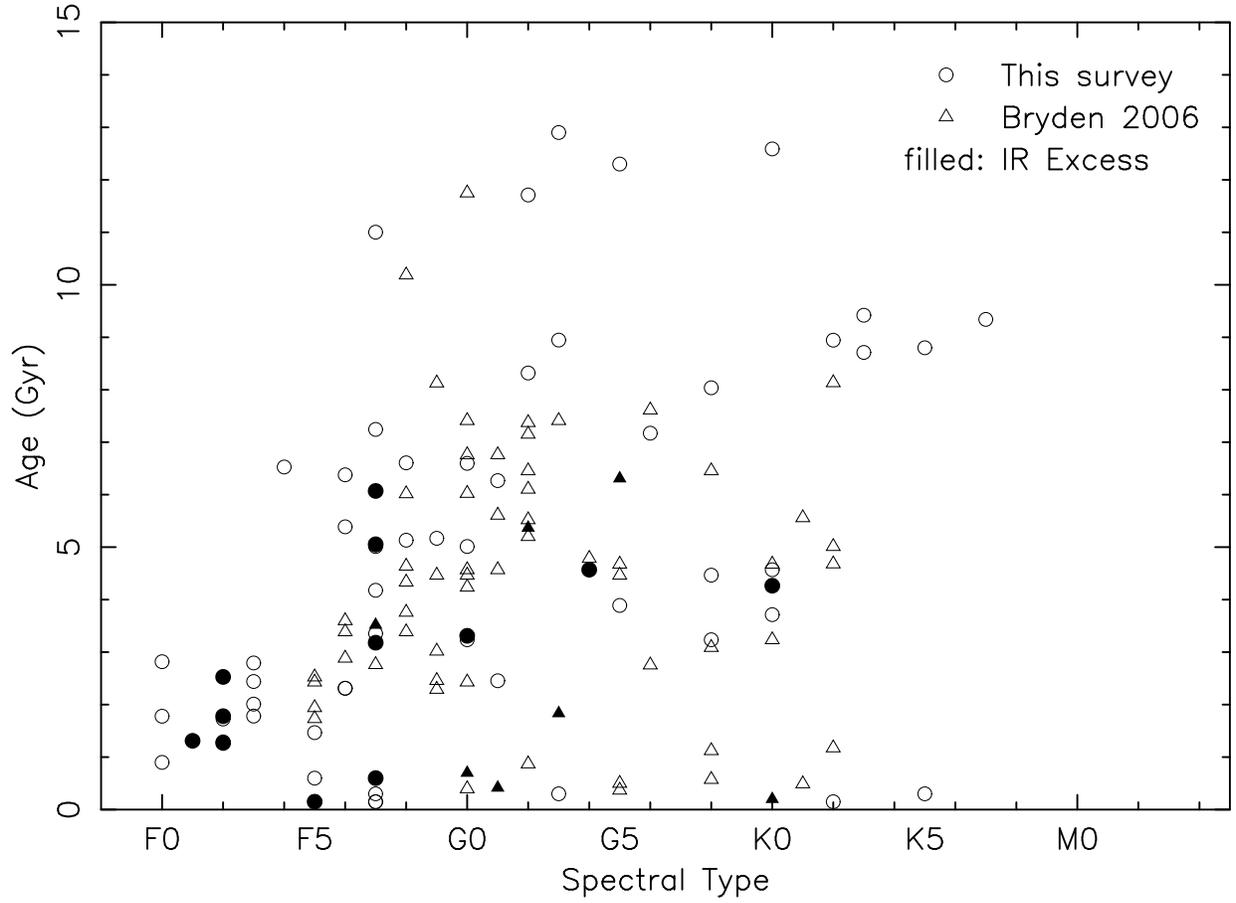} 
\end{center}
\caption{Stellar age as a function of spectral type for stars with
known ages. The stars from this survey are marked with circles, while
those from \citet{Bryden06a} are marked as triangles.
In both cases, stars with IR excess are marked as filled symbols.}
\label{sptypage}
\end{figure}

A possible interpretation of the trend with spectral type is that it
simply reflects the known correlation with stellar age. Earlier spectral type stars tend to be younger. Figure~\ref{sptypage} combines information on spectral type and age into a single plot for stars in this survey and those of \citet{Bryden06a}. The trends previously identified are apparent -
an upper limit to the ages of stars with excesses (filled symbols) of
about 6 Gyr and a tendency for earlier type stars to have excess more
frequently than later types. While the earliest type stars (F0-F3) are clearly younger on average, there is no clear evidence within the bulk of the sample that higher mass stars have more frequent excess because of they have younger ages. The formal correlation of excess with spectral type is even
stronger than the correlation with age (correlation coefficients
are -0.20$\pm$0.08. and -0.15$\pm$0.08 for spectral type and age
respectively), further suggesting that spectral type is an independent
indicator for IR excess. Unfortunately many of the latest type stars lack reliable age indicators, making it difficult to make any stronger conclusions.

\subsubsection{Comparison with Early Type Stars}

The detection rate of \70um excess around A stars is 33$\pm$4\% \citep{Su06},
more than twice that for the stars considered in this paper (13$\pm$3\%).
However, the A star and FGK star samples differ in both
mass and age. 
We first consider the possibility that the different detection rates simply reflect
an age evolution, rather than a spectral type dependence.
For example, the youngest FGK stars have a detection rate somewhat
higher than that within the sample as a whole:
considering only systems with ages of 0.1-1 Gyr (and including the stars
from \citet{Bryden06a}), 5 out of 19 young FGK stars 
have excess \70um emission (= 26$\pm$12\%).
Similarly, the \70um excess frequency among the A stars 
drops with stellar age down to just 21$\pm$6\% 
for A stars 0.3 - 1 Gyr old \citep{Su06}.
It is important to note, however, that 
many of the \citet{Su06} observations 
are less sensitive than those presented here,
relative to the stellar photosphere.
Thus, their A star detection rate should be regarded as a lower limit.
Although the FGK and A star samples have stellar age as the most important
correlating factor for IR excess,
we cannot rule out a
weaker but still important dependence of IR excess on some factor
related to stellar mass such as luminosity or disk mass.


\subsubsection{Comparison with Late Type Stars}

Combining the observations presented here with those of
\citet{Bryden06a} and \citet{Gautier06}, we have a total sample of 61
K1-M6 stars with no evidence of excess emission at \70um. Even
considering only those stars whose photospheres are detected at \70um
with S/N$>$3 (42 of the 61 stars), this lack of excess detections
is $>$3-$\sigma$ inconsistent with the $\sim$15\% detection rate
around F and G type stars. As implied by Eqn (3), the contrast of dust
relative to photosphere is, however, poorer for cooler stars which emit more of
their energy in the infrared than hotter stars. The average upper
limit to $\ld$ for the 16 stars K1 or later with detected photospheres
but no excesses in the SIM/TPF sample ($S/N (70 \mu {\rm m})>3$ and
$\chi_{70}<3$) is $\ld<9 \times 10^{-6}$ compared with the average upper
limit for 51 hotter stars with detected photospheres but no excesses,
$\ld<4\times 10^{-6}$. Thus, one explanation for the lack of excesses
around later-type stars is simply that the effective observational
limits are a factor of two higher for the cooler stars. While
observational selection effects make detection of IR excess around
late type stars more difficult, the strength of this trend suggests
that other explanations are needed. 

Another ambiguity in interpreting the correlation of excess with 
spectral type results from our limited knowledge of the location of
the dust. 
If dust around later type stars is very distant from its central star,
it will be too cool for detection at \70um.
Figure~\ref{Tbytype} shows how the dust
temperature varies as a function of spectral type for stars with
excess from both this sample and from \citet{Bryden06a}. We can only
derive a dust temperature in the limited number of cases where we have
a measured excess at both 24 and \70um ({\it solid
points}). Otherwise, only upper limits can be 
obtained (Table~\ref{excesstable}). For unresolved disk observations, the dust location cannot be determined without some knowledge of the underlying dust emission properties. Smaller dust with low emissivity can be just as hot as larger grains closer to the central star. Lines of constant orbital radius are shown in Figure~\ref{Tbytype} under the assumption of either large blackbody grains ($\epsilon =1$) or of small grains with emissivity = 0.01. Although the observed temperatures range from 80 to 170 K, they are all more or less consistent with emission from similar orbital locations: 10 AU for $\epsilon = 1$ or 100 AU for $\epsilon = 0.01$. By implication, one would expect that dust around later K stars might have typical temperatures of $\sim$50 K, ideal for detection at \70um, though none were detected. 

Additional information on the location of the dust comes from IRS observations of F, G, and K stars with excesses \citep{Beichman06irs} which reveal that in almost all cases the inner boundary of the emitting region occurs at or interior to 10 AU. A theoretical basis for this preferred location of a radial distance of a few AU comes from the suggestion that the water-ice sublimation distance, or the ``snowline'' where the temperature falls below 170 K, should mark the onset of the region of giant planet formation and its remnants in the Kuiper Belt \citep{hayashi81,sasselov2000, garaud06}. Since the location of the snowline varies with stellar luminosity ($\propto L^{0.25}$), there is no reason to expect a more distant, hidden reservoir of material unsampled by our observations around cooler stars. It is, of course, important to verify this expectation with observations at longer wavelengths such as MIPS 160 $\mu$m and in the sub-millimeter. 
Within the \citet{Gautier06} sample, for example, none of the 20 M
stars examined at \160um show any excess emission, providing limits
on $\ld$ of $10^{-5}$-10$^{-3}$ for material at $\sim$50 AU.

If the lack of debris disks around cool stars is real, then the dearth of material might reflect different formation mechanisms and evolutional history for the belts of planetesimals around low mass stars. Dust-producing collisions within these belts, for example, may require planetesimal stirring by larger, gas-giant planets, whose frequency is thought to be lower for late-type stars \citep{laugh04}, Alternately, the lack of IR excess might instead indicate a change in the physics of the smallest orbiting bodies as later type stars are considered, such as the increased relative importance of stellar winds in clearing dust from the system \citep{plavchan05}. 

\section{Applicability to TPF}\label{imp}

The detection of other terrestrial planets is a long term goal for the
astronomical community \citep{mckee01}. NASA has spent considerable
funds over the past decade on technology development and mission
studies for a Terrestrial Planet Finder (TPF). One of the key
astro-engineering issues revealed by those studies is the level of
dust emission associated with target stars since exo-zodiacal emission
is potentially an important source of photon shot noise
\citep{beichman99a}. Thus, in addition to scientific interest, the
incidence and distribution of material in the habitable zones,
i.e. where planets might have surface temperatures consistent with the
presence of liquid water \citep{kasting1993}, of nearby stars is of considerable technical importance. 

\subsection{Effect of Exo-Zodiacal Dust on Planet Finding}

As discussed in Appendix \ref{appendix}, dust emission at the level of 10-20 times that of our own zodiacal cloud can impede planet searches (Figure~\ref{snrplot}) due to increased photon shot noise for either a coronagraph or an interferometer. Since this level is roughly 50-100 times less than that presently detectable with \spit, we can rule out only those stars with the most extreme zodiacal disks. Thus, HD~109085 and HD~69830 \citep{Beichman05irs} are unsuitable targets with strong excess shortward of \24um. However, the remaining stars in this sample and other samples pass the initial screening by having \24um excesses, if any, less than $\ld \simeq 10^{-4}$, corresponding to upper limits on $\mu_{EZ}\sim$ 500 \citep{Bryden06a}. Beyond these photometric constraints, IRS spectroscopy can push upper limits to factors of 2-3 lower than MIPS alone and can also identify stars with small grain emission at 10 $\mu$m \citep{Beichman06irs}. 

In a few cases listed in Table~\ref{TPFexcesstable} we can use the
blackbodies fitted to the emission from the 5 stars with data at both

24 and 70 $\micron$ (Table~\ref{excesstable}) to extrapolate the emission from this ``Kuiper Belt'' dust to the prime TPF-I wavelength of 10 $\micron$. The extrapolated emission is also given in units of $\ld$ for material emitting at 10 $\micron$ (cf.\ eq.~[\ref{ldeq}] and eq.~[2] of \citet{Beichman06irs}) relative to the solar system value of $10^{-7}$ \citep{backman93}. Emission from any of this material located within the primary beam of the TPF-I telescopes ($r<$5 AU for a star at 10 pc observed with 3~m apertures) would be a noise source as described in the Appendix. However, this population of ``cool'' or ``lukewarm'' grains would not be located within a TPF-C pixel centered on the $\sim$ 1 AU habitable zone and would not be a noise source at visible wavelengths.

Unfortunately, however, the present observations cannot rule out an additional population of hotter grains located closer to the star which would either emit at 10 $\micron$ or scatter in the visible. IRS observations in the 8-14 $\mu$m region reach levels of just 1,000 times the zodiacal level \citep{Beichman06irs}. It will take observations with nulling Interferometers such as the Keck and Large Binocular Telescope Interferometers which can spatially suppress the stellar component to measure directly the exo-zodiacal emission in the habitable zone at levels that could cause S/N or confusion problems for TPF. 

There is some cause for optimism, however. The ``luminosity function''
of disks inferred from a variety of \spit\ samples \citep{Bryden06a},
the rarity of extreme ``hot'' zodiacal disks in the sample reported
here and in other \spit\ papers \citep{Bryden06a, Beichman06irs}, and
the apparent decline in the number of stars with excesses as a function of age (Fig.~\ref{ages}) are all encouraging signs that the relatively clean example of our solar system may be the norm rather than the exception. The ring-like structures seen in a number of resolved Spitzer disks, e.g. Fomalhaut \citep{Stapelfeldt04} and $\epsilon$ Eri (Backman et al., in preparation) as well as in HST images \citep{Kalas05, Kalas06} suggest that although the regions interior to the rings may not be completely empty due to a variety of mechanisms capable of transporting material inward from the outer disk \citep[comets, PR drag, interactions with planets, etc.;][]{Holmes02}, these interior regions may have a quite low total optical depth, perhaps as low as the $\sim$20\% contribution inferred for material from Kuiper Belt material to the total amount seen at 1 AU in our solar system \citep{Landgraf02, Dermott04, moro05}. 

\section{Summary}\label{conclusions}

We have searched for circumstellar dust around a sample of 88 F-M
stars, by means of photometric measurements at \24um and \70um. We
detected all the stars at \24um with high S/N and more than 80\% of
the stars at \70um with S/N $> 3$. Uncertainties in the \spit\
calibration and in the extrapolation of stellar photospheres to far-IR
wavelengths limit our ability to detect IR excesses with 3-$\sigma$ confidence to $\sim$20\% and $\sim$50\% of the photospheric levels at 24 and \70um, respectively. 

At these levels we have detected 12 of 88 objects with significant
\70um excesses.
Combined with an earlier study \citep{Bryden06a},
we find an overall detection rate of $13 \pm 3$\% for mature cool stars.
Beyond the single previously known \24um excess within our sample,
we detect two objects with \70um excesses and definite but weak
\24um emission. Another two stars with \70um excesses have $2-\sigma$
hints of \24um excesses. These results build on 
the finding of \citet{Beichman06irs} that in many cases, objects with
\70um emission also had IRS spectra rising longward of 25~$\um$ to
meet the \70um excess. These objects are all consistent with a disk
architecture similar to our Kuiper Belt that is concentrated outside
5-10 AU. In this context we note that a number of the \70um sources
are slightly, but significantly extended at \70um. The detailed
discussion of these objects is deferred to a subsequent paper
\citep{Bryden06b}. The IR emission in these systems is different from
the exceptional object HD~69830, which shows a disk architecture much
more consistent with a massive asteroid belt \citep{Beichman05irs}.  

Cross-correlating the detections of IR excess with stellar parameters
we find no significant correlations in the incidence of excesses
metallicity, but do find weak correlations with both stellar age and
spectral type. The lack of correlation with metallicity contrasts
with the known correlation between planet detections and stellar
metallicity, and the expectation that higher metal content might result in a greater number of dust-producing planetesimals. 

One significant finding is that the incidence of debris disks
among mature stars is markedly lower for spectral types later than K0
than for earlier spectral types. Combining data from this survey, the
\citet{Bryden06a} F5-K5 survey, and the \citet{Gautier06} M star
survey suggests an incidence of disks of $15\pm$3\% for F0-K0 stars
and $0\pm$4\% for stars with types K2-M3. This lack of disks around later
spectral types may be due to selection effects, lower initial disk
mass, or different rates of dust creation or destruction.

The disks that we are detecting have typical \70um luminosities around
100 times that of the Kuiper Belt. If they also have inner asteroid
belts 100 times brighter than our own, however, we would still not be
able to detect this warm inner dust. The observed \70um excess systems
could all be scaled-up replicas of the solar system's dust disk architecture, 
differing only in overall magnitude. These systems could have planets, asteroids and Kuiper Belt Objects as in our own system, but simply with a temporarily greater amount of dust due to a recent collisional event. 
Further observations of the warmer inner dust are necessary to 
address this possibility. \spit/IRS is particularly promising in this
regard \citep{Beichman06irs} and is being pursued as part of a
follow-up effort for some of the stars in this program. 

\acknowledgments {This publication makes use of data products from the
Two-Micron All Sky Survey (2MASS), as well as from IPAC, SIMBAD,
VIZIER, and the ROE Debris Disks Database website. We gratefully
acknowledge the assistance of John Carpenter in reducing the IRAC data
reported in this paper and we thank Angelle Tanner and Kate Su for helpful discussions.
The {\it Spitzer Space Telescope} is operated by the Jet Propulsion Laboratory, California Institute of Technology, under NASA contract 1407. Development of MIPS was funded by NASA through the Jet Propulsion Laboratory, subcontract 960785. Some of the research described in this publication was carried out at the Jet Propulsion Laboratory, California Institute of Technology, under a
contract with the National Aeronautics and Space Administration.} 

\appendix
\section{Noise Due to Exo-Zodiacal Emission}\label{appendix}

In this section we make order of magnitude estimates of the impact of photon noise from exo-zodiacal emission on both visible light and mid-IR instruments (TPF-Coronagraph and TPF-Interferometer, respectively) designed to find neighboring planets. A detailed noise analysis of planet finding telescopes is beyond the scope of this paper and the reader is referred to other articles for more details \citep{beichman99b, brown05}.

\subsection{TPF-I, The Infrared Interferometer}

The use of a nulling interferometer to reject starlight and thereby 
reveal an orbiting planet dates to an article by \citet{Bracewell78} and has been further investigated through studies of more sophisticated configurations \citep{Angel97, Lay05}. For a cryogenic system operating in an orbit near 1 AU, the three dominant noise sources are \citep[Table~\ref{snr}]{beichman99b}: the stellar light that leaks past the interferometric null because of the finite diameter of the star, S$_{*,Leak}$; emission from the local zodiacal dust, S$_{LZ}$; and emission from the exo-zodiacal dust in the target star system that leaks past the interferometer, S$_{EZ,Leak}$ (see
Figure~\ref{schematic}). At short wavelengths ($<$8 $\micron$), the
stellar leak may dominate all other noise sources; longward of 20 
$\micron$ emission from a 35 K telescope will become important; and at
all wavelengths various systematic instrumental effects will be important. But over a broad range of wavelengths, the balance between S$_{*,Leak}$, S$_{LZ}$, S$_{EZ,Leak}$ controls the fundamental noise floor. Detector read noise and dark current can be ignored for broad band detection. 

In the background limit considered here, the total noise is given by
the square root of the sum of all the individual photon fluxes reaching the detector. Rather than evaluate the absolute signal-to-noise ratio, S/N, we consider here the ratio of the S/N in the presence of exo-zodiacal emission, SNR(EZ), to the S/N in the absence of such emission, SNR(0):

\begin{equation}
 \left.{{SNR(EZ)}\over{SNR(0)}} \right|_{IR}=
{
{\sqrt{ S_{*,Leak} +S_{LZ}}}
\over
{\sqrt{ S_{*,Leak} +S_{LZ} + S_{EZ,Leak}}}}
={{\sqrt{ 1 + { {S_{*,Leak}} \over {S_{LZ}} }}}
\over
{\sqrt{ 1+ { {S_{*,Leak}} \over {S_{LZ}}}
 + { {S_{EZ, Leak}} \over {S_{LZ}}} }}}
\label{snreq2}
\end{equation}

In the above, $S_{*,Leak}$ depends on the nulling configuration, the
wavelength of operation and the angular size of the star. Null depths
of $10^{-5}$ to $10^{-6}$ have been demonstrated in the laboratory
\citep{Martin03} and for the purposes of this illustration, it
suffices to take $S_{*,Leak}=10^{-5}F_{*}$. The emission from the
local zodiacal cloud, $S_{LZ}$, is very complex in detail
\citep{Kelsall98}, but can be parameterized for our purposes as
follows, $S_{LZ} = \tau_{LZ} B_\nu(255\, K)\Omega_{tel}$ where $B_\nu$
is the Planck function, $\tau_{LZ }$ is the vertical optical depth
looking out from the mid-ecliptic plane at 1 AU, and $\Omega_{tel}$ is
the diffraction limited solid angle of an individual telescope in the
interferometer. A typical value of the zodiacal cloud brightness
toward the ecliptic pole from our mid-plane location is 12 MJy
sr$^{-1}$ at 12 $\micron$ \citep{Kelsall98}. 

In the absence of more detailed information, the vertical optical depth of the exo-zodiacal dust in any system can be parameterized as a factor, $\mu_{EZ}$, times the Solar System's zodiacal dust. The emission from exo-zodiacal dust is then $S_{EZ}(r)= 2\mu_{EZ}\tau_{LZ}(r) B_\nu(T(r)) \Omega_{tel}$, where 
the factor of two accounts for the fact that in the exo-zodiacal case we are looking through the entire cloud and not from the vantage of the mid-plane as we do the local cloud. By analogy with the local zodiacal cloud \citep{Backman98} the vertical optical depth is assumed to fall off radially as $\tau_{LZ}(r)= \tau_{LZ, 1 AU}r_{AU}^{-0.3}$. We also take $T(r)=T_0r_{AU}^\beta$ as the equilibrium temperature for grains heated by stellar radiation and emitting in the infrared. Typical 1 AU values of ($T_0$, $\beta$) for large and small silicate grains are (255 K, -0.5) and (362 K, -0.4), respectively \citep{Draine84, backman93, Beichman06irs}. The large and small grain brightness distributions are normalized to yield the same value at 1 AU.

The effect of exo-zodiacal emission is modulated by the fringe pattern of the interferometer which attenuates the bright central portion of the exo-zodiacal disk. To account for this effect we incorporate the fringe pattern of a particular nulling scheme $\zeta(\theta,\phi)$ where $\theta$ and $\phi$ are the radial and azimuthal variables, respectively. In the simplified case of a face-on disk, the signal reaching the detector, S$_{EZ, Leak}$ is then given by the integral of S$_{EZ}$ over the fringe pattern and the telescope solid angle:
\begin{equation}
S_{EZ,Leak}(d) = \int_0^{2\pi}d\phi \int_0^{\theta_{max}} 
\mu_{EZ}\tau_{LZ}(\theta d) B_\nu(T(\theta d)) \zeta(\theta,\phi) \theta d\theta
\end{equation}
for a star at a distance $d$. We adopt the fringe pattern 
$\zeta(\theta,\phi)$ for the Dual Chopped Bracewell interfermoter \citep[DCB;][]{Lay04, Lay05} presently under study. Canceling out common factors, the stellar leak term in equation~(\ref{snreq2}) then becomes 
\begin{equation}
 \left.{ {S_{EZ, Leak}(d)} \over {S_{LZ}}} \right|_{IR} = 2 \mu_{EZ} (e^{
{{14388}\over {\lambda T_{LZ}}}}-1) { {1} \over {\Omega_{tel}} }
\int_0^{2\pi}d\phi \int_0^{\theta_{max}} { {\zeta(\theta,\phi) (\theta d)^{-0.3}} \over
{e^{14388/(\lambda T(\theta d))} -1} } \theta d\theta 
\label{snreq4}
\end{equation}

To evaluate equation~(\ref{snreq4}), we adopt a diffraction limited
beam size of \break $\theta_{max} = 0.6\lambda/D =0.5^{\prime\prime}$ for a $D=$3 m telescope at 12 $\micron$. For a solar-type star at $d=$10 pc, the ratio of the exo-zodiacal contribution to that from the Solar System's own dust (eq.~\ref{snreq4}) is 0.06 $\mu_{EZ}$ or 0.24 $\mu_{EZ}$, for large and small grains respectively. Warmer, smaller grains fill more of the beam of the individual telescopes than the cooler, larger (blackbody) grains and thus contribute more noise. With this information in hand, Figure~\ref{snrplot} shows the variation of S/N as a function of exo-zodiacal brightness, $\mu_{EZ}$,
for two grain sizes. When the exo-zodiacal surface density $\mu_{EZ}$ is 10 times that of our solar system, corresponding to a 20-fold brightness increase, the S/N is reduced by a factor of $\sim$2, necessitating an increase in integration time by a factor of $\sim$4 to recover the original S/N. It is interesting to note the importance of grain size on this effect; the emission from the large grains is more centrally peaked and thus more effectively attenuated by the nulling interferometer than for the smaller grains which remain warm at quite large distances from the star. Since at least a few hours of integration time is needed to detect an Earth in the presence of a $\mu_{EZ}=1$ cloud, and days to carry out a spectroscopic program 
\citep{Beichman98, Lay05}, it is clear that studying systems with
$\mu_{EZ}>10-20$ will be difficult. 

\subsection{TPF-C, the Visible Light Coronagraph}

\begin{figure}
\begin{center}
\includegraphics[width=8in]{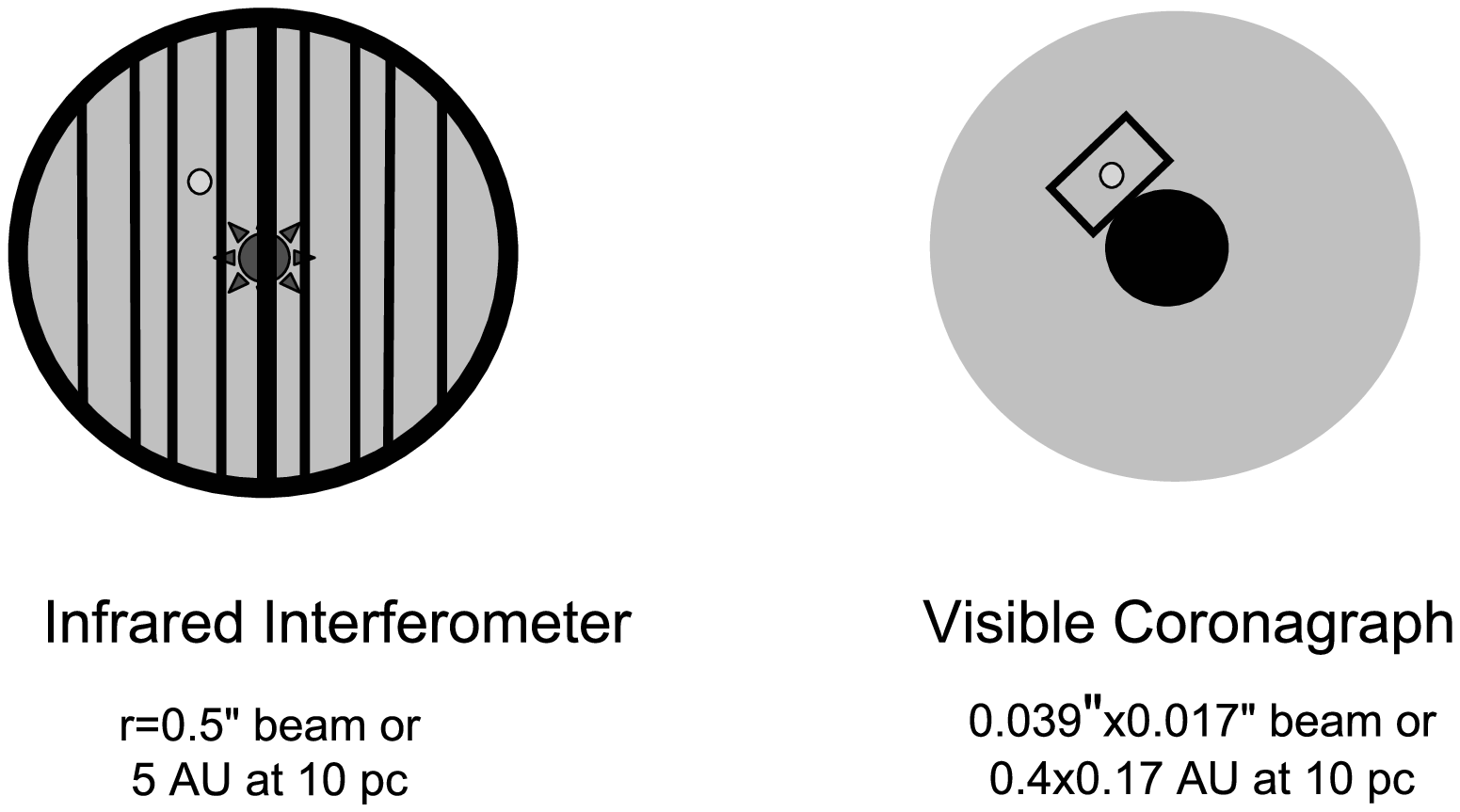}
\end{center}
\caption{The interferometer ({\it left}) takes in all emission from a face-on
exo-zodiacal disk that fits within the diffraction-limited beam of a single telescope. The intensity is, however, attenuated by the interferometric fringe pattern shown as vertical black bars. The coronagraph ({\it right}) takes in exo-zodiacal light only within the area of a single diffraction-limited pixel; the values shown here are appropriate for a 3.5x8 m telescope under consideration for TPF-C.} 
\label{schematic}
\end{figure}

A similar analysis can be applied to an assessment of the effects
of exo-zodiacal emission at visible wavelengths. There are some
important differences however. First, the coronagraph takes in only
the exo-zodiacal light from the immediate vicinity of the planet, not
from the entire exo-zodiacal cloud (Fig.~\ref{schematic}, {\it right side}). Second, the signal from an Earth ($S_p$), the residual starlight after a 10$^{-10}$ rejection ratio, and the local and exo-zodiacal signals are all more evenly balanced. Detector noise becomes a serious issue at medium spectral resolution ($\sim$75), but can be ignored in the broadband case. The analog of equation~(\ref{snreq2}) for the coronagraph becomes:
\begin{equation}
\left. { {SNR(EZ)}\over{SNR(0)}}\right|_{Vis}=
{\sqrt{1+ {{S_{*,residual}} \over {S_{LZ}}} + {{S_p}\over{S_{LZ}}} }
\over \sqrt{1+ {{S_{*,residual}} \over {S_{LZ}}} + {{S_p}\over{S_{LZ}}}
+ {{S_{EZ}}\over{S_{LZ}}} }}
\end{equation}

\begin{figure}
\begin{center}
\includegraphics[width=4.7in,angle=-90]{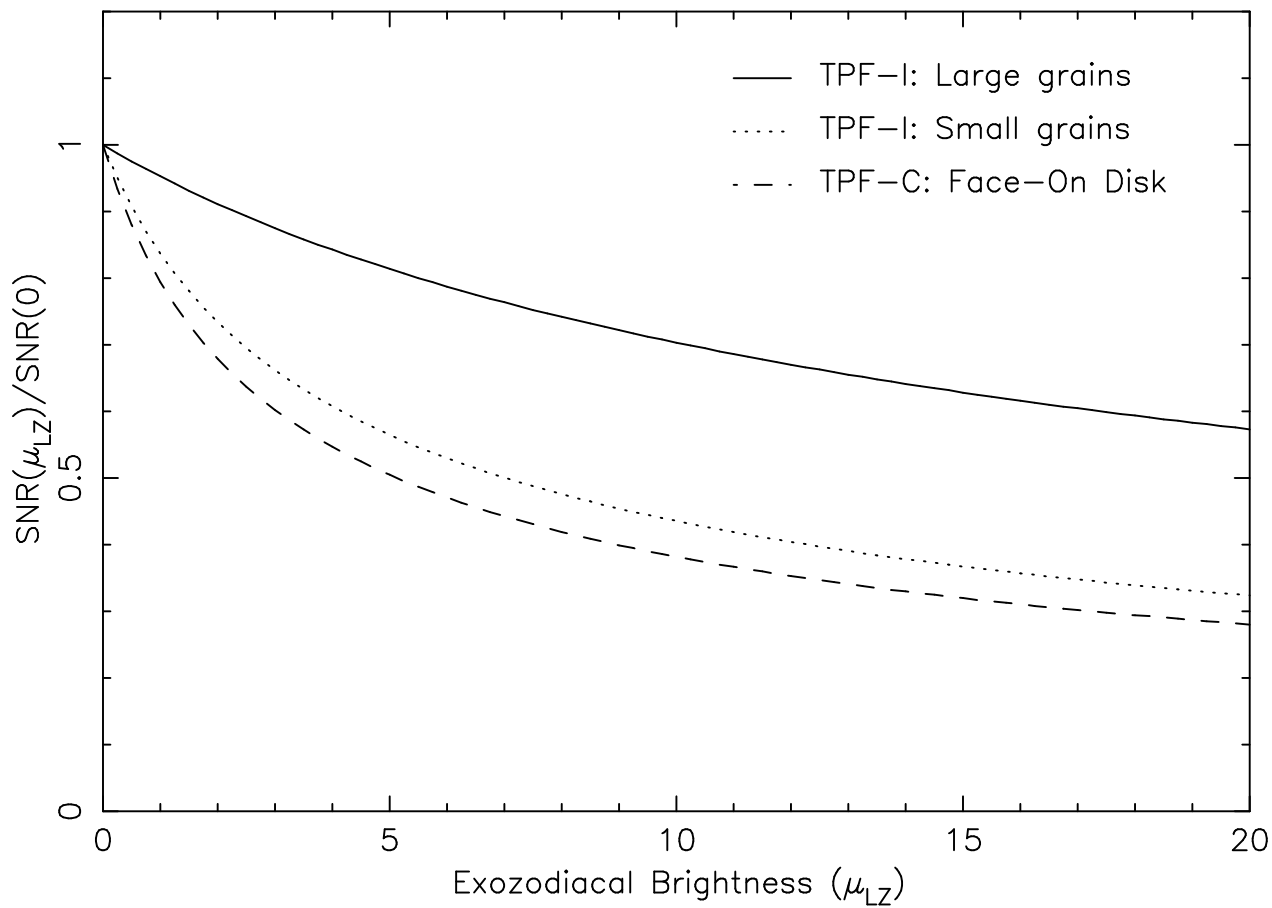} 
\end{center}
\caption{The effect of exo-zodiacal dust emission on TPF signal-to-noise ratio.
The horizontal axis gives the vertical optical depth of the exo-zodiacal disk normalized to that of the Solar System ($\mu_{LZ}$). Note that a value of $\mu_{EZ}=1$ in a target system corresponds to twice the emission we see from our zodiacal cloud, e.g. using COBE or IRAS, since we view our cloud from its midplane. The upper two curves show the falloff in relative S/N for the interferometer as the amount of exo-zodiacal emission increases, with large and small grain sizes considered separately (solid and dotted lines respectively). The lower curve (dashed line) shows a similar trend for a coronagraph viewing a face-on disk. In each case, the signal-to-noise ratio is shown relative to observations of a system with no dust emission.}
\label{snrplot}
\end{figure}

Since the local and exo-zodiacal emission enter the system through
exactly the same solid angle, $\Omega_{tel}$, the ${ {S_{EZ}} \over
{S_{LZ}}}$ term simplifies to 2$\mu_{EZ}$. For a planet 25 mag fainter
than a V=4.5 mag solar twin at 10 pc, and assuming a local zodiacal
brightness of 0.1 MJy sr$^{-1}$ at 0.55 $\micron$
\citep[Table~\ref{snr};][]{Bernstein02}, we can evaluate the variation
in S/N as a function of $\mu_{EZ}$. Figure~\ref{snrplot} shows the decrease in S/N as the exo-zodiacal emission increases in the case of a face-on disk; an edge-on disk will increase the surface brightness and resultant noise. As with the interferometer, the effect of zodiacal emission in the target system is to lower the S/N by a factor of 2$\sim$3 at $\mu_{EZ}=10.$ 

The {\it relative} effect of the exo-zodiacal emission is somewhat more pronounced for the TPF-C than for the TPF-I because the interferometer is dominated by the strong {\it local} zodiacal background until very bright exo-zodiacal levels are observed. The intrinsic background level within the visible-light coronagraph is low (by assumption of an excellent 10$^{-10}$ rejection ratio) so that the exo-zodiacal emission more quickly plays a significant role in setting the system noise. 

A more detailed examination of the effects of the exo-zodiacal emission on the detectability of planets using TPF-C and TPF-I would yield absolute, not relative, sensitivity levels including the effects of disk inclination and confusion by structures, e.g. wakes and gaps, within the zodiacal cloud. These questions lie beyond the scope of this paper.


\clearpage
\pagestyle{empty}
\begin{deluxetable}{l|llll|ccl|ccccc|ccc}
\tabletypesize{\scriptsize}
\tablewidth{0pt} \rotate \tablecaption{Basic Data\label{basictable}}
\scriptsize
\tablehead{
Star            &   HIP &   GJ  &   HR  &   other   &   Spectral    &   V   &  \hspace{0.07in} K           &    &       &   Age (Gyr)   &     &       &       &   [Fe/H]  &       \\
            &       &       &       &   name    &   Type    &   (mag)
&   (mag)           &   Mo/W/Average$^{a}$   &   Min &   Max &  \# est. &   References  &   Average &   $\sigma$    &   References  }
\startdata
GL 436$^{b}$      &   57087   &   436 &       &       &   M2.5    &   10.67   &   6.07            &   $-$ &   $-$ &   $-$ &   0   &   $-$ &   $-$ &   $-$ &   $-$ \\
GL 908          &   117473  &   908 &       &   BR Psc  &   M1  &   8.98    &   5.04            &   $-$ &   $-$ &   $-$ &   0   &   $-$ &   $-$ &   $-$ &   $-$ \\
HD 739          &   950 &   3013    &   35  &   $\theta$ Scl    &   F4V &   5.24    &   4.13$^{d}$  &   2.80    &   2.40    &   3.19    &   2   &   I,N &   -0.13   &   0.07    &   CS,E,I,M,N,T    \\
HD 4391         &   3583    &   1021    &   209 &       &   G5IV    &   5.80    &   4.30$^{d}$  &   12.30   &   $-$ &   $-$ &   1   &   N   &   -0.17   &   0.07    &   E,N,RP,T    \\
HD 4813         &   3909    &   37  &   235 &   19 Cet  &   F7IV-V  &   5.17    &   4.02$^{d}$  &   5.04    &   2.35    &   9.63    &   7   &   C,I,L,La,N,P    &   -0.16   &   0.09    &   CS,C,E,I,L,La,M,N,P,T   \\
HD 10360$^{c}$  &       &   66B &   486 &       &   K2V &   5.76    &   3.56$^{d}$  &   0.15    &   $-$ &   $-$ &   1   &   Mo  &   -0.23   &   0.03    &   E,N,RP,T,V  \\
HD 16895            &   12777   &   107A    &   799 &   $\theta$ Per    &   F7V &   4.10    &   2.98$^{e}$      &   5.01    &   2.50    &   7.94    &   5   &   W,C,L,N,P   &   -0.08   &   0.09    &   CS,C,E,L,M,N,P,T,V  \\
HD 20794            &   15510   &   139 &   1008    &   e Eri   &   G8V &   4.26    &   2.52$^{e}$      &   $-$ &   $-$ &   $-$ &   0   &   $-$ &   -0.32   &   0.09    &   CS,E,I,N,P,RP,T,V   \\
HD 22001            &       &   143.2A  &   1083    &   $\kappa$ Ret    &   F5IV-V  &   4.71    &   3.94$^{d}$  &   0.60    &   0.60    &   9.38    &   5   &   Mo,La,N &   -0.13   &   0.07    &   CS,La,M,N,T \\
HD 23249            &   17378   &   150 &   1136    &   $\delta$ Eri    &   K0IV    &   3.52    &   1.45$^{e}$      &   12.59   &   $-$ &   $-$ &   1   &   P   &   0.02    &   0.11    &   CS,P,RP,V   \\
HD 23754            &   17651   &   155 &   1173    &   27 Eri  &   F3/F5V  &   4.22    &   3.35$^{d}$  &   2.01    &   1.40    &   3.02    &   4   &   F,I,M,N &   0.05    &   0.08    &   CS,F,I,M,N,T    \\
HD 25998            &   19335   &   161 &   1278    &   50 Per  &   F7V &   5.52    &   4.28$^{d}$  &   0.60    &   0.60    &   5.14    &   4   &   Mo,C,L,N    &   -0.01   &   0.10    &   CS,C,L,M,N,T    \\
HD 28343            &   20917   &   169 &       &       &   K7V     &   8.30    &   4.88            &   $-$ &   $-$ &   $-$ &   0   &   $-$ &   $-$ &   $-$ &   $-$ \\
HD 32147            &   23311   &   183 &   1614    &       &   K3V     &   6.22    &   3.71$^{d}$  &   $-$ &   $-$ &   $-$ &   0   &   $-$ &   0.16    &   0.14    &   CS,E,I,P,T,V    \\
HD 36395            &   25878   &   205 &       &       &   M1.5V   &   7.97    &   3.86$^{e}$      &   $-$ &   $-$ &   $-$ &   0   &   $-$ &   0.60    &   $-$ &   CS  \\
HD 38392$^{c}$  &       &   216B    &   1982    &   $\gamma$ Lep B  &   K2V &   6.15    &   4.13$^{d}$  &   8.94    &   8.75    &   9.14    &   3   &   La  &   -0.05   &   0.09    &   CS,E,La,M,N,RP,T    \\
HD 38858            &   27435   &   1085    &   2007    &       &   G4V &   5.97    &   4.41$^{d}$  &   4.57    &   3.19    &   12.20   &   3   &   W,I,N   &   -0.25   &   0.01    &   I,N,T,V \\
HD 39587            &   27913   &   222 &   2047    &   54 Ori  &   G0V &   4.39    &   2.97$^{e}$      &   6.60    &   0.10    &   10.70   &   5   &   B,C,L,N,P   &   -0.07   &   0.07    &   CS,C,E,L,M,N,P,T,V  \\
HD 40136            &   28103   &   225 &   2085    &   $\eta$ Lep  &   F1V     &   3.71    &   2.90$^{e}$      &   1.31    &   1.22    &   1.41    &   3   &   I,N,P   &   -0.16   &   0.06    &   CS,I,M,N,P  \\
HD 46588            &   32439   &   240 &   2401    &       &   F8V &   5.44    &   4.14$^{d}$  &   5.13    &   4.27    &   6.20    &   4   &   B,F,I,N &   -0.22   &   0.07    &   F,I,M,N,RP  \\
HD 48682            &   32480   &   245 &   2483    &   56 Aur  &   G0V     &   5.24    &   4.13$^{d}$  &   3.31    &   3.31    &   8.91    &   5   &   W,B,M,N,P   &   0.07    &   0.08    &   CS,E,M,N,P,T,V  \\
HD 50281$^{c}$  &   32984   &   250A    &       &       &   K3V &   6.58    &   4.11$^{d}$  &   9.42    &   9.02    &   9.82    &   3   &   La  &   0.06    &   0.07    &   CS,La,M,T,V \\
HD 53706$^{c}$  &   34069   &   264.1B  &   2668    &       &   K0V &   6.83    &   4.94            &   $-$ &   $-$ &   $-$ &   0   &   $-$ &   -0.24   &   0.05    &   CS,E,N,RP,T,V   \\
HD 55892            &   34834   &   268 &   2740    &   QW Pup  &   F0IV    &   4.49    &   3.71$^{d,e}$  &   1.78    &   1.40    &   2.16    &   2   &   L,N &   -0.30   &   0.10    &   CS,L,M,N    \\
HD 62644            &   37606   &       &   2998    &   GJ 284  &   G6IV    &   5.04    &   3.12$^{d}$  &   7.17    &   3.41    &   14.13   &   3   &   I,P,R   &   -0.09   &   0.21    &   CS,I,P,R,T  \\
HD 63077            &   37853   &   288A    &   3018    &   171 Pup &   G0V &   5.36    &   3.75$^{d}$  &   5.01    &   5.01    &   14.50   &   4   &   W,C,I,N &   -0.79   &   0.11    &   CS,C,E,I,M,N,P,T    \\
HD 67228            &   39780   &       &   3176    &   $\mu$ Cnc   &   G2IV    &   5.30    &   3.83$^{e}$      &   8.32    &   5.50    &   8.32    &   4   &   W,F,I,N &   0.11    &   0.06    &   CS,F,I,N,T,V    \\
HD 68146            &   40035   &   297.2A  &   3202    &   18 Pup  &   F7V &   5.53    &   4.35$^{d}$  &   4.18    &   2.92    &   5.19    &   4   &   C,L,M,N &   -0.13   &   0.10    &   CS,C,E,L,M,N,T  \\
HD 71243            &   40702   &   305 &   3318    &   $\alpha$ Cha    &   F5V &   4.05    &   3.15$^{d}$  &   1.47    &   1.40    &   1.53    &   2   &   F,N &   0.07    &   0.02    &   F,M,N   \\
HD 72673            &   41926   &   309 &   3384    &       &   K0V     &   6.38    &   4.44$^{d}$  &   4.57    &   $-$ &   $-$ &   1   &   W   &   -0.36   &   0.06    &   CS,E,I,N,RP,T,V \\
HD 76653            &   43797   &   3519    &   3570    &       &   F6V     &   5.70    &   4.56$^{d}$  &   2.31    &   2.10    &   2.52    &   2   &   I,N &   -0.04   &   0.07    &   I,M,N   \\
HD 76932            &   44075   &   3523    &   3578    &       &   F7/F8IV/V   &   5.80    &   4.36$^{d}$  &   11.00   &   9.29    &   12.50   &   4   &   C,F,I,N &   -0.84   &   0.12    &   CS,C,E,F,I,L,M,N,T  \\
HD 78366            &   44897   &   334 &   3625    &       &   F9V     &   5.95    &   4.55            &   5.17    &   3.84    &   6.50    &   2   &   I,N &   0.02    &   0.09    &   E,I,M,N,V   \\
HD 79211$^{c}$  &   120005  &   338B    &       &       &   K2  &   7.70    &   4.14$^{e}$      &   $-$ &   $-$ &   $-$ &   0   &   $-$ &   $-$ &   $-$ &   $-$ \\
HD 81937            &   46733   &   3559    &   3757    &   h UMa   &   F0IV    &   3.65    &   2.82$^{d,e}$  &   0.90    &   $-$ &   $-$ &   1   &   N   &   0.06    &   $-$ &   N   \\
HD 81997            &   46509   &   348A    &   3759    &   31 Hya  &   F6V &   4.59    &   3.56$^{d}$  &   6.38    &   1.94    &   9.43    &   5   &   La,M,N  &   0.00    &   0.01    &   E,La,M,N    \\
HD 85512            &   48331   &       &       &   GJ 370  &   K5V     &   7.67    &   4.72            &   0.30    &   $-$ &   $-$ &   1   &   Mo  &   $-$ &   $-$ &   $-$ \\
HD 89449            &   50564   &   388 &   4054    &   40 Leo  &   F6IV    &   4.78    &   3.65$^{d,e}$  &   2.31    &   1.64    &   3.40    &   4   &   F,I,M,N &   0.02    &   0.08    &   CS,F,I,M,N,T    \\
HD 90089            &   51502   &   392 &   4084    &       &   F2V &   5.25    &   4.27$^{d}$  &   1.78    &   1.50    &   2.06    &   2   &   I,N &   -0.28   &   0.10    &   I,M,N,T \\
HD 90589            &   50954   &   391 &   4102    &   I Car   &   F2IV    &   3.99    &   3.12$^{e}$      &   1.73    &   0.40    &   3.33    &   3   &   I,M,N   &   0.01    &   0.14    &   I,M,N   \\
HD 91324            &   51523   &   397 &   4134    &       &   F6V     &   4.89    &   3.58$^{d}$  &   5.39    &   4.28    &   7.94    &   4   &   L,M,N,P &   -0.54   &   0.35    &   CS,L,M,N,P,T    \\
HD 100623           &   56452   &   432A    &   4458    &       &   K0V &   5.96    &   4.02$^{d}$  &   3.72    &   3.72    &   10.08   &   4   &   W,La    &   -0.38   &   0.10    &   E,La,M,N,RP,T,V \\
HD 102365           &   57443   &   442A    &   4523    &       &   G5V &   4.89    &   3.31$^{e}$      &   8.95    &   6.12    &   10.08   &   4   &   I,La    &   -0.36   &   0.14    &   CS,E,I,La,N,P,RP,T,V    \\
HD 103095           &   57939   &   451A    &   4550    &   CF UMa  &   G8V &   6.42    &   4.37$^{e}$      &   3.24    &   3.24    &   5.40    &   2   &   W,B &   -1.35   &   0.02    &   CS,E,N,P,T,V    \\
HD 105211           &   59072   &   455 &   4616    &   $\eta$ Cru  &   F2V &   4.14    &   3.20$^{d}$  &   2.53    &   1.30    &   3.99    &   3   &   F,I,N   &   -0.37   &   0.18    &   I,M,N   \\
HD 105452           &   59199   &   455 &   4623    &   $\alpha$ Crv    &   F0IV/V  &   4.02    &   3.17$^{e}$      &   2.82    &   $-$ &   $-$ &   1   &   P   &   -0.43   &   0.26    &   CS,N,P  \\
HD 109085           &   61174   &   471 &   4775    &   $\eta$ Crv  &   F2V &   4.30    &   3.54$^{e}$      &   1.27    &   0.95    &   1.56    &   3   &   I,M,N   &   -0.05   &   0.04    &   I,M,N   \\
HD 129502           &   71957   &   9491    &   5487    &   $\mu$ Vir   &   F2V &   3.87    &   2.90$^{e}$      &   1.29    &   0.71    &   1.71    &   4   &   F,I,M,N &   0.03    &   0.09    &   F,I,M,N \\
HD 131977$^{c}$  &   73184   &   570A    &   5568    &       &   K4V &   5.72    &   3.15$^{e}$      &   $-$ &   $-$ &   $-$ &   0   &   $-$ &   0.07    &   0.07    &   CS,P,RP,T,V \\
HD 132254           &   73100   &   3880    &   5581    &       &   F7V &   5.63    &   4.41            &   3.35    &   2.16    &   3.96    &   6   &   C,F,I,L,M,N &   0.01    &   0.05    &   CS,C,F,L,M,N,RP,T   \\
HD 136352  S         &   75181   &   582 &   5699    &       &   G2V     &   5.65    &   4.16$^{d}$  &   11.71   &   7.52    &   15.90   &   2   &   I,N &   -0.36   &   0.08    &   CS,E,I,N,P,RP,T,V   \\
HD 139664           &   76829   &   594 &   5825    &   g Lup   &   F5IV-V  &   4.64    &   3.80$^{d}$  &   0.15    &   0.15    &   9.30    &   6   &   Mo,I,La,N   &   -0.15   &   0.09    &   I,La,M,N    \\
HD 142267           &   77801   &   3924    &   5911    &   39 Ser  &   G0V &   6.07    &   4.53$^{d}$  &   3.24    &   3.24    &   13.50   &   3   &   W,N,P   &   -0.34   &   0.14    &   CS,E,M,N,P,T,V  \\
HD 147513$^{b}$      &   80337   &   620.1A  &   6094    &       &   G5V &   5.37    &   3.93$^{e}$      &   0.30    &   0.30    &   8.50    &   2   &   Mo,N    &   0.02    &   0.11    &   CS,M,N,RP,T,V   \\
HD 151288           &   82003   &   638 &       &       &   K7V     &   8.10    &   4.71            &   $-$ &   $-$ &   $-$ &   0   &   $-$ &   $-$ &   $-$ &   $-$ \\
HD 154363           &   83591   &   653 &       &       &   K5V     &   7.70    &   4.73            &   $-$ &   $-$ &   $-$ &   0   &   $-$ &   $-$ &   $-$ &   $-$ \\
HD 156026           &   84478   &   664 &       &   36 Oph C    &   K5V     &   6.33    &   3.47$^{d}$  &   8.80    &   8.64    &   8.96    &   3   &   La  &   -0.16   &   0.07    &   CS,La,P,T   \\
HD 157881           &   85295   &   673 &       &       &   K7V     &   7.54    &   4.14$^{e}$      &   9.34    &   $-$ &   $-$ &   1   &   La  &   0.00    &   0.35    &   CS,I    \\
HD 158633           &   85235   &   675 &   6518    &       &   K0V     &   6.44    &   4.52$^{d}$  &   4.27    &   $-$ &   $-$ &   1   &   W   &   -0.43   &   0.08    &   E,I,M,N,V   \\
HD 160032           &   86486   &   686 &   6569    &   $\lambda$ Ara   &   F3IV    &   4.76    &   3.83$^{d}$  &   2.44    &   1.85    &   3.30    &   4   &   F,I,M,N &   -0.29   &   0.06    &   CS,F,I,M,N,T    \\
HD 164259           &   88175   &   699 &   6710    &   $\zeta$ Ser &   F3V     &   4.62    &   3.64$^{d}$  &   1.78    &   1.34    &   2.06    &   5   &   F,I,L,M,N   &   -0.11   &   0.06    &   CS,F,I,L,M,N    \\
HD 165499           &   89042   &   705 &   6761    &   \emph{i} Pav    &   G1V &   5.47    &   4.13$^{d}$  &   6.27    &   2.65    &   10.80   &   4   &   I,L,M,N &   -0.14   &   0.07    &   CS,E,I,L,M,N,RP \\
HD 172051           &   91438   &   722 &   6998    &       &   G5V     &   5.85    &   4.23$^{d}$  &   3.89    &   1.54    &   3.89    &   2   &   W,I &   -0.28   &   0.03    &   E,I,N,RP,V  \\
HD 177565           &   93858   &   744 &   7232    &       &   G8V &   6.15    &   4.54$^{d}$  &   8.04    &   5.01    &   13.20   &   3   &   I,N,R   &   0.05    &   0.02    &   CS,E,I,N,R,RP,T,V   \\
HD 180617           &   94761   &   752A    &       &       &   M2.5    &   9.12    &   4.67$^{e}$      &   $-$ &   $-$ &   $-$ &   0   &   $-$ &   $-$ &   $-$ &   $-$ \\
HD 182488           &   95319   &   758 &   7368    &       &   G8V     &   6.37    &   4.49$^{d}$  &   4.47    &   4.47    &   10.52   &   2   &   W,I &   0.11    &   0.08    &   E,I,M,N,RP,V    \\
HD 185395           &   96441   &   765A    &   7469    &   $\theta$ Cyg    &   F4V &   4.49    &   3.54$^{d}$  &   6.53    &   1.50    &   9.24    &   5   &   La,N,P  &   -0.04   &   0.08    &   CS,La,M,N,P,T,V \\
HD 187691           &   97675   &   768.1A  &   7560    &   \emph{o} Aql    &   F8V &   5.12    &   3.90$^{d}$  &   6.61    &   3.70    &   9.00    &   6   &   W,B,C,L,M,N &   0.09    &   0.04    &   CS,C,L,M,N,T,V  \\
HD 189245           &   98470   &   773 &   7631    &       &   F7V &   5.65    &   4.48$^{d}$  &   0.15    &   0.15    &   5.20    &   3   &   Mo,I,N  &   -0.26   &   0.07    &   I,M,N   \\
HD 190406           &   98819   &   779 &   7672    &   15 Sge  &   G1V     &   5.80    &   4.39$^{d}$  &   2.45    &   2.45    &   8.80    &   4   &   W,B,M,N &   -0.05   &   0.06    &   CS,E,M,N,RP,V   \\
HD 191849           &   99701   &   784 &       &       &   M0V     &   7.97    &   4.28$^{e}$      &   $-$ &   $-$ &   $-$ &   0   &   $-$ &   $-$ &   $-$ &   $-$ \\
HD 192310           &   99825   &   785 &   7722    &       &   K0V &   5.73    &   3.50$^{d}$  &   8.71    &   $-$ &   $-$ &   1   &   La  &   -0.03   &   0.10    &   CS,E,I,N,P,T,V  \\
HD 196877           &   102186  &   798 &       &       &   K7V     &   8.83    &   5.47            &   $-$ &   $-$ &   $-$ &   0   &   $-$ &   $-$ &   $-$ &   $-$ \\
HD 198149           &   102422  &   807 &   7957    &   $\eta$ Cep  &   K0IV    &   3.41    &   1.28$^{e}$      &   $-$ &   $-$ &   $-$ &   0   &   $-$ &   -0.16   &   0.05    &   CS  \\
HD 199260           &   103389  &   811 &   8013    &       &   F7V &   5.70    &   4.48$^{d}$  &   3.18    &   2.90    &   3.46    &   2   &   I,N &   -0.20   &   0.12    &   I,M,N   \\
HD 213845           &   111449  &   863 &   8592    &   $\upsilon$ Aqr  &   F7V &   5.21    &   4.33$^{d,e}$  &   0.15    &   0.15    &   2.32    &   4   &   Mo,I,M,N    &   0.02    &   0.10    &   I,M,N,T \\
HD 215648           &   112447  &   872A    &   8665    &   $\xi$ Peg   &   F7V &   4.20    &   2.96$^{d}$  &   7.24    &   2.24    &   7.24    &   5   &   W,L,M,N,P   &   -0.27   &   0.09    &   CS,L,M,N,P,T,V  \\
HD 217357           &   113576  &   884 &       &       &   K5  &   7.88    &   4.48            &   $-$ &   $-$ &   $-$ &   0   &   $-$ &   $-$ &   $-$ &   $-$ \\
HD 219482           &   114948  &   1282    &   8843    &       &   F7V     &   5.64    &   4.44$^{d}$  &   6.07    &   5.60    &   6.54    &   2   &   F,N &   -0.16   &   0.06    &   F,M,N   \\
HD 219623           &   114924  &   4324    &   8853    &       &   F7V     &   5.58    &   4.31$^{d}$  &   5.06    &   4.60    &   5.50    &   4   &   C,L,M,N &   -0.05   &   0.09    &   CS,C,E,L,M,N,T  \\
HD 222237           &   116745  &   902 &       &       &   K3V     &   7.09    &   4.58            &   $-$ &   $-$ &   $-$ &   0   &   $-$ &   -0.16   &   0.14    &   E,M,N,T,V   \\
HD 265866           &   33226   &   251 &       &       &   M3.5
&   9.89    &   5.28            &   $-$ &   $-$ &   $-$ &   0   &
$-$ &   $-$ &   $-$ &   $-$ \\  \hline Companions
\\  \hline
HD 10360J           &   7751    &   66  &       &   p Eri   &   K0V &   5.07    &   3.51$^{d}$  &   $-$ &   $-$ &   $-$ &   0   &   $-$ &   -0.28   &   0.00    &   E,T \\
HD 38393            &   27072   &   216A    &   1983    &   $\gamma$ Lep    &   F7V &   3.59    &   2.42$^{e}$      &   0.30    &   0.30    &   9.51    &   6   &   Mo,C,L,La   &   -0.08   &   0.05    &   CS,C,E,L,La,M,N,P,T \\
HD 50281B           &       &   250B    &       &       &   M2  &   10.10   &   5.72            &   $-$ &   $-$ &   $-$ &   0   &   $-$ &   $-$ &   $-$ &   $-$ \\
HD 53705            &   34065   &   264.1A  &   2667    &       &   G3V  &   5.56    &   4.04$^{d}$  &   12.90   &   $-$ &   $-$ &   1   &   N   &   -0.27   &   0.06    &   CS,E,N,RP,T,V   \\
HD 79210            &   45343   &   338A    &       &       &   K7  &   7.64    &   3.99            &   $-$ &   $-$ &   $-$ &   0   &   $-$ &   $-$ &   $-$ &   $-$ \\
HD 131976           &   73182   &   570B    &       &       &   M1V &   8.01    &   3.90$^{e}$      &   $-$ &   $-$ &   $-$ &   0   &   $-$ &   $-$ &   $-$ &   $-$ \\
\enddata
\tablenotetext{a}{Age from Montes' paper, then from Wright's paper if
available, otherwise an average of literature values}
\tablenotetext{b}{Known planet-bearing star} 
\tablenotetext{c}{Star has a wide binary companion that was included in the
survey, as listed at the bottom of the table} 
\tablenotetext{d}{Star has one or more bad 2MASS values (err $>$ 20$\%$)} 
\tablenotetext{e}{Star has JHK values from Johnson or other literature}
\tablecomments{Spectral types from SIMBAD.  Visual magnitudes are as
quoted in SIMBAD, typically from the Hipparcos satellite;
K magnitues are from 2MASS unless otherwise noted.} 
\tablerefs{(B)  \citet{barry88};
(B)  \citet{barry88};
(CS) \citet{cayrel96,cayrel01};
(C)  \citet{chen01};
(E)  \citet{eggen98};
(F)  \citet{feltzing01};
(I)  \citet{Ibukiyama02};
(L)  \citet{Lambert04};
(La) \citet{Lachaume99};
(M)  \citet{Marsakov88,Marsakov95};
(Mo)  \citet{Montes01};
(N)  \citet{Nordstrom04};
(P)   \citet{Perrin77};
(R)   \citet{Randich99};
(RP)  \citet{Rocha-Pinto98};
(T)  \citet{Taylor02};
(W)  \citet{Wright04};
(V)  \citet{Valenti05}}
\end{deluxetable}

\begin{deluxetable}{l|cccc|cccccc}
\tabletypesize{\scriptsize}
\tablewidth{0pt} \rotate \tablecaption{Measured and predicted flux
densities at 24 and 70 $\mu$m (in mJy)\label{mipstable}}
\tablehead{
            &       &   24 $\mu$m   &       &       &               &       &   70 $\mu$m   &       &       &               \\
Star            &   \emph{F}$_{\nu,MIPS}$   &   \emph{F}$_{\nu,*}$
&   \emph{F}$_{\nu,MIPS}$/\emph{F}$_{\nu,*}$    &
$\chi_{24}^{a}$ &   \emph{F}$_{\nu,MIPS}$           &
\emph{F}$_{\nu,*}$  &   \emph{F}$_{\nu,MIPS}$/\emph{F}$_{\nu,*}$
&   S/N &   $\chi_{70}^{b}$ &
\emph{L}$_{dust}$/\emph{L}$_{*}^{c}$          } \startdata
GL 436$^{}$  &   38.7    &   31.3    &   1.24    &   6.0 &   4.0 $\pm$   2.2 &   3.5 &   1.1 &   1.9 &   0.2 &   $<$     5.4  $\times$ 10$^{-5}$ \\
GL 908          &   96.3    &   99.2    &   0.97    &   -0.7    &   12.0    $\pm$   3.3 &   10.9    &   1.1 &   4.3 &   0.3 &   $<$     2.2  $\times$ 10$^{-5}$ \\
HD 739          &   159.1   &   142.5   &   1.12    &   1.4 &   16.2    $\pm$   3.8 &   16.2    &   1.0 &   5.6 &   0.0 &   $<$     3.3  $\times$ 10$^{-6}$ \\
HD 4391         &   141.3   &   150.7   &   0.94    &   -1.6    &   19.5    $\pm$   3.5 &   17.2    &   1.1 &   8.4 &   0.7 &   $<$     3.9  $\times$ 10$^{-6}$ \\
HD 4813         &   191.8   &   207.3   &   0.93    &   -0.9    &   21.2    $\pm$   4.6 &   23.5    &   0.9 &   7.2 &   -0.5    &   $<$     2.7  $\times$ 10$^{-6}$ \\
HD 10360            &   247.7   &   283.6   &   0.87    &   -1.6    &   23.1    $\pm$   6.1 &   32.4    &   0.7 &   6.3 &   -1.5    &   $<$     4.6  $\times$ 10$^{-6}$ \\
HD 16895            &   492.7   &   458.7   &   1.07    &   1.9 &   51.2    $\pm$   9.7 &   51.9    &   1.0 &   8.8 &   -0.1    &   $<$     2.4  $\times$ 10$^{-6}$ \\
HD 20794            &   737.2   &   770.8   &   0.96    &   -1.1    &   94.3    $\pm$   13.6    &   87.9    &   1.1 &   28.7    &   0.5 &   $<$     1.2  $\times$ 10$^{-6}$ \\
HD 22001            &   233.7   &   205.8   &   1.14    &   1.7 &   25.2    $\pm$   4.8 &   23.3    &   1.1 &   7.8 &   0.4 &   $<$     2.7  $\times$ 10$^{-6}$ \\
HD 23249            &   2039.1  &   1825.2  &   1.12    &   2.9 &   207.0   $\pm$   32.1    &   206.5   &   1.0 &   24.5    &   0.0 &   $<$     1.5  $\times$ 10$^{-6}$ \\
HD 23754            &   383.3   &   368.0   &   1.04    &   0.5 &   46.6    $\pm$   7.1 &   41.8    &   1.1 &   13.8    &   0.7 &   $<$     1.5  $\times$ 10$^{-6}$ \\
HD 25998$^{d,e}$      &   147.1   &   128.5   &   1.14    &   3.6 &   61.9    $\pm$   5.7 &   14.5    &   4.3 &   11.7    &   8.3 &       2.7  $\times$ 10$^{-5}$ \\
HD 28343            &   85.4    &   98.4    &   0.87    &   -3.3    &   -9.5    $\pm$   14.8    &   11.4    &   -0.8    &   -0.6    &   -1.4    &   $<$     8.8  $\times$ 10$^{-5}$ \\
HD 32147            &   224.3   &   240.6   &   0.93    &   -0.8    &   23.1    $\pm$   6.3 &   27.5    &   0.8 &   4.9 &   -0.7    &   $<$     7.6  $\times$ 10$^{-6}$ \\
HD 36395            &   250.7   &   288.9   &   0.87    &   -3.3    &   26.7    $\pm$   17.6    &   30.2    &   0.9 &   1.6 &   -0.2    &   $<$     4.9  $\times$ 10$^{-5}$ \\
HD 38392$^{}$  &   203.7   &   156.9   &   1.30    &   3.7 &   18.8    $\pm$   8.2 &   17.7    &   1.1 &   2.4 &   0.1 &   $<$     1.8  $\times$ 10$^{-5}$ \\
HD 38858$^{e,f}$      &   131.3   &   131.3   &   1.00    &   0.0 &   153.7   $\pm$   9.8 &   15.0    &   10.3    &   16.0    &   14.1    &       1.0  $\times$ 10$^{-4}$ \\
HD 39587            &   483.6   &   488.5   &   0.99    &   -0.3    &   35.3    $\pm$   12.1    &   55.4    &   0.6 &   4.0 &   -1.7    &   $<$     4.0  $\times$ 10$^{-6}$ \\
HD 40136$^{d,e}$      &   553.3   &   490.5   &   1.13    &   3.2 &   90.7    $\pm$   9.4 &   55.6    &   1.6 &   21.3    &   3.7 &       4.0  $\times$ 10$^{-6}$ \\
HD 46588            &   150.8   &   161.8   &   0.93    &   -1.7    &   14.6    $\pm$   3.7 &   18.3    &   0.8 &   6.1 &   -1.0    &   $<$     3.0  $\times$ 10$^{-6}$ \\
HD 48682$^{e,f}$      &   188.4   &   188.8   &   1.00    &   0.0    &   256.9   $\pm$   6.8 &   21.3    &   12.1    &   46.3    &   36.8    &       1.1  $\times$ 10$^{-4}$ \\
HD 50281            &   175.8   &   174.5   &   1.01    &   0.2 &   17.1    $\pm$   5.4 &   19.9    &   0.9 &   3.8 &   -0.5    &   $<$     10.0     $\times$ 10$^{-6}$ \\
HD 53706            &   72.5    &   73.4    &   0.99    &   -0.3    &   7.0 $\pm$   2.7 &   8.3 &   0.8 &   3.0 &   -0.5    &   $<$     1.0  $\times$ 10$^{-5}$ \\
HD 55892            &   263.8   &   243.0   &   1.09    &   2.1 &   25.5    $\pm$   5.7 &   27.6    &   0.9 &   6.6 &   -0.4    &   $<$     2.2  $\times$ 10$^{-6}$ \\
HD 62644            &   370.6   &   369.0   &   1.00    &   0.1 &   46.2    $\pm$   8.5 &   41.9    &   1.1 &   8.2 &   0.5 &   $<$     4.0  $\times$ 10$^{-6}$ \\
HD 63077            &   226.4   &   240.9   &   0.94    &   -1.5    &   15.6    $\pm$   8.0 &   27.6    &   0.6 &   2.3 &   -1.5    &   $<$     6.2  $\times$ 10$^{-6}$ \\
HD 67228            &   215.6   &   208.3   &   1.04    &   0.9 &   14.4    $\pm$   5.0 &   23.5    &   0.6 &   4.0 &   -1.8    &   $<$     4.1  $\times$ 10$^{-6}$ \\
HD 68146            &   132.6   &   132.5   &   1.00    &   0.0 &   18.3    $\pm$   3.6 &   15.1    &   1.2 &   6.6 &   0.9 &   $<$     4.0  $\times$ 10$^{-6}$ \\
HD 71243            &   451.1   &   434.5   &   1.04    &   0.5 &   52.9    $\pm$   8.2 &   49.3    &   1.1 &   14.5    &   0.4 &   $<$     1.5  $\times$ 10$^{-6}$ \\
HD 72673            &   119.2   &   132.8   &   0.90    &   -2.6    &   9.2 $\pm$   3.0 &   15.2    &   0.6 &   4.6 &   -2.0    &   $<$     4.8  $\times$ 10$^{-6}$ \\
HD 76653            &   112.3   &   109.1   &   1.03    &   0.7 &   33.9    $\pm$   10.5    &   12.4    &   2.7 &   3.3 &   2.1 &   $<$     1.7  $\times$ 10$^{-5}$ \\
HD 76932            &   130.2   &   140.7   &   0.93    &   -1.9    &   15.2    $\pm$   3.8 &   16.1    &   0.9 &   5.3 &   -0.2    &   $<$     3.9  $\times$ 10$^{-6}$ \\
HD 78366            &   108.3   &   115.3   &   0.94    &   -1.5    &   16.8    $\pm$   4.2 &   13.1    &   1.3 &   4.6 &   0.9 &   $<$     6.8  $\times$ 10$^{-6}$ \\
HD 79211            &   195.4   &   187.1   &   1.04    &   1.1 &   18.0    $\pm$   4.0 &   21.4    &   0.8 &   7.4 &   -0.8    &   $<$     4.7  $\times$ 10$^{-6}$ \\
HD 81937            &   563.9   &   558.4   &   1.01    &   0.2 &   70.0    $\pm$   10.4    &   63.3    &   1.1 &   16.6    &   0.6 &   $<$     1.0  $\times$ 10$^{-6}$ \\
HD 81997            &   287.8   &   264.4   &   1.09    &   1.1 &   28.2    $\pm$   5.7 &   30.0    &   0.9 &   8.0 &   -0.3    &   $<$     2.4  $\times$ 10$^{-6}$ \\
HD 85512            &   95.9    &   110.9   &   0.86    &   -3.4    &   11.0    $\pm$   3.0 &   12.8    &   0.9 &   4.7 &   -0.6    &   $<$     1.0  $\times$ 10$^{-5}$ \\
HD 89449            &   240.8   &   255.1   &   0.94    &   -1.4    &   24.6    $\pm$   5.3 &   28.9    &   0.8 &   8.2 &   -0.8    &   $<$     2.1  $\times$ 10$^{-6}$ \\
HD 90089$^{e}$      &   147.0   &   146.7   &   1.00    &   0.0 &   38.2    $\pm$   3.6 &   16.7    &   2.3 &   14.9    &   6.0 &       8.5  $\times$ 10$^{-6}$ \\
HD 90589            &   422.9   &   409.8   &   1.03    &   0.8 &   53.7    $\pm$   8.5 &   46.4    &   1.2 &   11.0    &   0.9 &   $<$     1.8  $\times$ 10$^{-6}$ \\
HD 91324            &   263.2   &   273.9   &   0.96    &   -0.5    &   50.5    $\pm$   9.2 &   31.2    &   1.6 &   6.4 &   2.1 &   $<$     5.3  $\times$ 10$^{-6}$ \\
HD 100623           &   183.8   &   193.4   &   0.95    &   -1.2    &   21.8    $\pm$   4.1 &   22.1    &   1.0 &   9.0 &   -0.1    &   $<$     4.0  $\times$ 10$^{-6}$ \\
HD 102365           &   353.9   &   353.4   &   1.00    &   0.0 &   34.0    $\pm$   8.5 &   40.2    &   0.8 &   5.7 &   -0.7    &   $<$     4.1  $\times$ 10$^{-6}$ \\
HD 103095           &   128.5   &   133.8   &   0.96    &   -1.0    &   9.3 $\pm$   3.1 &   15.4    &   0.6 &   4.4 &   -1.9    &   $<$     4.4  $\times$ 10$^{-6}$ \\
HD 105211$^{e,f}$      &   367.9   &   363.4   &   1.01    &   0.2 &   473.7   $\pm$   19.8    &   41.4    &   11.4    &   25.2    &   21.8    &       6.9  $\times$ 10$^{-5}$ \\
HD 105452           &   396.9   &   409.1   &   0.97    &   -0.7    &   43.3    $\pm$   8.4 &   46.6    &   0.9 &   9.3 &   -0.4    &   $<$     1.6  $\times$ 10$^{-6}$ \\
HD 109085$^{d,e,f}$  &   589.2   &   296.8   &   1.99    &   24.6    &   198.2   $\pm$   6.8 &   33.5    &   5.9 &   42.7    &   24.0    &       3.3  $\times$ 10$^{-5}$ \\
HD 129502           &   536.0   &   521.1   &   1.03    &   0.7 &   51.8    $\pm$   9.7 &   59.1    &   0.9 &   13.1    &   -0.8    &   $<$     1.2  $\times$ 10$^{-6}$ \\
HD 131977           &   427.8   &   397.4   &   1.08    &   1.9 &   37.6    $\pm$   7.4 &   45.3    &   0.8 &   12.6    &   -1.0    &   $<$     3.2  $\times$ 10$^{-6}$ \\
HD 132254           &   127.5   &   131.5   &   0.97    &   -0.8    &   23.6    $\pm$   3.4 &   14.9    &   1.6 &   9.1 &   2.5 &   $<$     3.8  $\times$ 10$^{-6}$ \\
HD 136352           &   169.8   &   175.1   &   0.97    &   -0.8    &   17.5    $\pm$   5.1 &   19.9    &   0.9 &   4.2 &   -0.5    &   $<$     5.5  $\times$ 10$^{-6}$ \\
HD 139664$^{e,f}$      &   275.9   &   251.9   &   1.10    &   1.2 &   503.7   $\pm$   9.2 &   28.6    &   17.6    &   62.2    &   51.9    &       1.3  $\times$ 10$^{-4}$ \\
HD 142267           &   111.4   &   107.5   &   1.04    &   0.9 &   9.5 $\pm$   2.7 &   12.2    &   0.8 &   4.7 &   -1.0    &   $<$     4.2  $\times$ 10$^{-6}$ \\
HD 147513           &   201.1   &   195.4   &   1.03    &   0.7 &   17.3    $\pm$   12.1    &   22.1    &   0.8 &   1.5 &   -0.4    &   $<$     1.4  $\times$ 10$^{-5}$ \\
HD 151288           &   102.7   &   115.9   &   0.89    &   -2.8    &   13.1    $\pm$   2.7 &   13.4    &   1.0 &   7.4 &   -0.1    &   $<$     9.0  $\times$ 10$^{-6}$ \\
HD 154363           &   95.3    &   110.5   &   0.86    &   -3.4    &   9.7 $\pm$   4.2 &   12.7    &   0.8 &   2.6 &   -0.7    &   $<$     1.6  $\times$ 10$^{-5}$ \\
HD 156026           &   304.2   &   306.9   &   0.99    &   -0.1    &   30.6    $\pm$   10.9    &   35.2    &   0.9 &   3.2 &   -0.4    &   $<$     1.5  $\times$ 10$^{-5}$ \\
HD 157881           &   171.3   &   192.8   &   0.89    &   -2.8    &   13.1    $\pm$   5.5 &   22.2    &   0.6 &   3.0 &   -1.7    &   $<$     1.3  $\times$ 10$^{-5}$ \\
HD 158633$^{e}$      &   112.3   &   125.1   &   0.90    &   -2.6    &   56.7    $\pm$   3.6 &   14.4    &   3.9 &   19.9    &   11.8    &       4.1  $\times$ 10$^{-5}$ \\
HD 160032           &   243.8   &   229.4   &   1.06    &   0.8 &   44.3    $\pm$   6.1 &   26.1    &   1.7 &   9.4 &   3.0 &   $<$     3.3  $\times$ 10$^{-6}$ \\
HD 164259           &   246.4   &   226.0   &   1.09    &   1.1 &   8.3 $\pm$   7.0 &   25.6    &   0.3 &   1.4 &   -2.5    &   $<$     4.1  $\times$ 10$^{-6}$ \\
HD 165499           &   174.4   &   182.1   &   0.96    &   -0.5    &   15.9    $\pm$   4.5 &   20.6    &   0.8 &   4.9 &   -1.1    &   $<$     4.1  $\times$ 10$^{-6}$ \\
HD 172051           &   150.2   &   147.4   &   1.02    &   0.5 &   25.5    $\pm$   11.5    &   16.7    &   1.5 &   2.3 &   0.8 &   $<$     1.9  $\times$ 10$^{-5}$ \\
HD 177565           &   110.9   &   109.8   &   1.01    &   0.2 &   16.4    $\pm$   4.9 &   12.4    &   1.3 &   3.6 &   0.8 &   $<$     1.2  $\times$ 10$^{-5}$ \\
HD 180617$^{}$  &   138.4   &   111.6   &   1.24    &   6.0 &   5.9 $\pm$   13.4    &   12.7    &   0.5 &   0.4 &   -0.5    &   $<$     9.6  $\times$ 10$^{-5}$ \\
HD 182488           &   109.4   &   108.7   &   1.01    &   0.2 &   2.2 $\pm$   10.7    &   12.3    &   0.2 &   0.2 &   -0.9    &   $<$     2.7  $\times$ 10$^{-5}$ \\
HD 185395           &   284.9   &   254.5   &   1.12    &   1.5 &   34.0    $\pm$   5.3 &   28.8    &   1.2 &   11.1    &   1.0 &   $<$     2.0  $\times$ 10$^{-6}$ \\
HD 187691           &   212.0   &   211.5   &   1.00    &   0.0 &   21.0    $\pm$   6.8 &   23.9    &   0.9 &   3.6 &   -0.4    &   $<$     5.6  $\times$ 10$^{-6}$ \\
HD 189245           &   120.8   &   116.5   &   1.04    &   0.9 &   6.6 $\pm$   3.9 &   13.3    &   0.5 &   1.9 &   -1.7    &   $<$     5.6  $\times$ 10$^{-6}$ \\
HD 190406           &   126.6   &   133.0   &   0.95    &   -1.2    &   22.6    $\pm$   5.1 &   15.1    &   1.5 &   5.0 &   1.5 &   $<$     7.9  $\times$ 10$^{-6}$ \\
HD 191849           &   179.0   &   192.8   &   0.93    &   -1.8    &   26.9    $\pm$   4.1 &   21.2    &   1.3 &   10.3    &   1.4 &   $<$     9.8  $\times$ 10$^{-6}$ \\
HD 192310           &   250.4   &   236.4   &   1.06    &   0.7 &   19.1    $\pm$   5.6 &   26.8    &   0.7 &   5.0 &   -1.4    &   $<$     6.3  $\times$ 10$^{-6}$ \\
HD 196877           &   54.1    &   59.4    &   0.91    &   -2.2    &   2.2 $\pm$   2.2 &   6.8 &   0.3 &   1.1 &   -2.1    &   $<$     2.0  $\times$ 10$^{-5}$ \\
HD 198149           &   2444.6  &   2418.4  &   1.01    &   0.3 &   254.6   $\pm$   42.0    &   276.7   &   0.9 &   38.5    &   -0.5    &   $<$     8.6  $\times$ 10$^{-7}$ \\
HD 199260$^{d,e}$      &   120.1   &   108.5   &   1.11    &   2.7 &   42.8    $\pm$   4.1 &   12.3    &   3.5 &   11.7    &   7.4 &       2.1  $\times$ 10$^{-5}$ \\
HD 213845           &   158.7   &   144.7   &   1.10    &   1.2 &   23.9    $\pm$   4.0 &   16.4    &   1.5 &   7.5 &   1.9 &   $<$     4.3  $\times$ 10$^{-6}$ \\
HD 215648           &   505.7   &   516.7   &   0.98    &   -0.3    &   47.4    $\pm$   9.6 &   58.8    &   0.8 &   12.7    &   -1.2    &   $<$     1.4  $\times$ 10$^{-6}$ \\
HD 217357           &   133.4   &   139.3   &   0.96    &   -1.1    &   18.6    $\pm$   3.5 &   16.1    &   1.2 &   7.5 &   0.7 &   $<$     8.6  $\times$ 10$^{-6}$ \\
HD 219482$^{d,e}$      &   140.9   &   131.0   &   1.08    &   1.9 &   65.4    $\pm$   3.7 &   14.9    &   4.4 &   22.5    &   13.8    &       2.8  $\times$ 10$^{-5}$ \\
HD 219623$^{e}$      &   144.6   &   139.7   &   1.04    &   0.9 &   48.0    $\pm$   3.8 &   15.8    &   3.0 &   15.9    &   8.4 &       1.7  $\times$ 10$^{-5}$ \\
HD 222237           &   108.4   &   112.0   &   0.97    &   -0.8    &   20.5    $\pm$   2.8 &   12.8    &   1.6 &   9.9 &   2.7 &   $<$     7.1  $\times$ 10$^{-6}$ \\
HD 265866           &   84.8    &   95.2    &   0.89    &   -2.7
&   $-$         &   9.9 &   $-$ &   $-$ &   $-$ &   $-$         \\
\hline Companions
\\  \hline
HD 10360J           &   241.0   &   337.7   &   0.71    &   -3.6    &   23.8    $\pm$   6.8 &   38.2    &   0.6 &   6.5 &   -2.1    &   $<$     3.5  $\times$ 10$^{-6}$ \\
HD 38393            &   765.4   &   761.6   &   1.00    &   0.1 &   57.3    $\pm$   15.1    &   86.4    &   0.7 &   7.3 &   -1.9    &   $<$     2.0  $\times$ 10$^{-6}$ \\
HD 50281B$^{}$  &   55.3    &   43.7    &   1.26    &   6.6 &   10.1    $\pm$   4.6 &   5.0 &   2.0 &   2.2 &   1.1 &   $<$     8.3  $\times$ 10$^{-5}$ \\
HD 53705            &   166.4   &   192.1   &   0.87    &   -3.3    &   17.3    $\pm$   4.0 &   21.9    &   0.8 &   7.3 &   -1.1    &   $<$     3.0  $\times$ 10$^{-6}$ \\
HD 79210            &   191.4   &   260.3   &   0.74    &   -6.6    &   21.4    $\pm$   5.1 &   29.8    &   0.7 &   8.8 &   -1.6    &   $<$     5.6  $\times$ 10$^{-6}$ \\
HD 131976           &   274.4   &   246.8   &   1.11    &   2.8 &   15.6    $\pm$   5.2 &   28.2    &   0.6 &   5.2 &   -2.4    &   $<$     9.2  $\times$ 10$^{-6}$ \\
\enddata
\tablenotetext{a}{Significance of \24um  excess (eq.~[\ref{chi24eq}])} 
\tablenotetext{b}{Significance of \70um excess (eq.~[\ref{chi70eq}])} 
\tablenotetext{c}{Minimum $\ld$ from \70um emission (eq.~[\ref{ldeq}])}
\tablenotetext{d}{Star with excess \24um emission}
\tablenotetext{e}{Star with excess \70um emission}
\tablenotetext{f}{Star with resolved \70um emission}
\end{deluxetable}

\normalsize
\begin{deluxetable}{cccccc} 
\tablecaption{IRAC Observations of M Stars$^a$\label{irac}}
\tablehead{
& Spectral & F$_\nu$(3.55 $\mu$m) & F$_\nu$(4.49 $\mu$m) &
F$_\nu$(5.73 $\mu$m)& F$_\nu$(7.87  $\mu$m) \\
Star & Type & (Jy) & (Jy) & (Jy) & (Jy) }
\startdata
GL908 	& M1 		& 3.29 & 2.22 & 1.43 & 0.82\\
HD~36395 	& M1.5V 	& 7.64 & 5.62 & 3.74 & 2.22\\
HD~191849 	& K7/M0 	& 6.46 & 4.12 & 2.70 & 1.58\\
HD~265866 	& M3.5 	& 2.75 & 1.83 & 1.22 & 0.70\\
\enddata
\tablenotetext{a}{{Flux density uncertainties are dominated by calibration uncertainties, typically 5\% for the IRAC bands.} }
\end{deluxetable}

\begin{deluxetable}{r|cccc} 
\tablecaption{Dust Emission at 24 and 70 $\mu$m (in mJy)\label{excesstable}} 
\tablehead{Star&\emph{F}$_\nu(24)_{dust}$ & \emph{F}$_\nu(70)_{dust}^a$ & 
   \emph{T}$_{dust}^b$ & $\ld^c$ } 
\startdata 
HD 25998&18.6$\pm$ 5.9&51.1 $\pm$ 9.6 & 96$\pm$5 &4.5 $\times10^{-5}$\\
HD 38858$^d$&$<15\, (3\sigma)$&193 $\pm$ 25 &$<70$&12 $\times10^{-5}$\\
HD 40136&63$\pm$ 22&37.9 $\pm$ 11& 165$_{-20}^{+35}$&1.9 $\times10^{-5}$\\
HD 48682$^d$&$<22\, (3\sigma)$&290 $\pm$ 38 &$<68$&1.1 $\times10^{-4}$\\
HD 90089&$<18\, (3\sigma)$&23.2 $\pm$ 5.0 &$<120$& 0.8$\times10^{-5}$\\
HD 105211$^d$& $<45\, (3\sigma)$&521 $\pm$ 73&$<70$&7.3 $\times10^{-5}$\\
HD 109085$^d$& 292$\pm$ 24&212 $\pm$ 28& 150$\pm$10 &15 $\times10^{-5}$\\
HD 139664$^d$& $<75\, (3\sigma)$&523 $\pm$ 77&$<78$&12 $\times10^{-5}$\\
HD 158633& $<13\, (3\sigma)$&45.8 $\pm$ 7.8&$<90$&3.5 $\times10^{-5}$\\
HD 199260& 11.6$\pm$ 5.6&32.9 $\pm$ 6.4&94$\pm$5&3.3 $\times10^{-5}$\\
HD 219482& 9.8$\pm$ 5.6&54.5 $\pm$ 9.0&81$\pm$3&3.6 $\times10^{-5}$\\
HD 219623& $<17\, (3\sigma)$&34.8 $\pm$ 6.5&$<104$& 1.6 $\times10^{-5}$\\
\enddata 
\tablenotetext{a}{Dust fluxes at \70um have been color corrected by 8\%.}
\tablenotetext{b}{Blackbody temperature based on either the 24 to 70
$\mu$m flux density ratio or the 70 $\mu$m flux density plus a 3$\sigma$
upper limit at 24 $\mu$m.}
\tablenotetext{c}{If only \70um data are available, $\ld$ is 
from Eqn~\ref{ldeq}. If \24um and \70um data are available, $\ld$
obtained from Eqn~\ref{ldeq2470}.} 
\tablenotetext{d}{For the resolved sources, the emission is fit with
an extended Gaussian profile, resulting in measured dust fluxes 
$\sim$10\% higher than from the standard aperture photometry.} 
\end{deluxetable}

\begin{deluxetable}{l|cc} 
\tablecaption{Parameters for TPF S/N Calculations \label{snr}}
\tablehead{Parameter&TPF-C&TPF-I}
\startdata
Wavelength 		& 0.55 $\mu$m 	&12  $\mu$m \\
Telescope 		& 3.5x8m		&Four, 3m  \\
			& 			&  on 75 m baseline\\
Beam Half Width 	& 39x17 mas 	&500 mas \\
Beam Area, $\Omega_{tel}$&$5\times10^{-14}$&$1.8\times 10^{-11}$\\
\hline
Local Zodiacal Emission ($I_{LZ}$) 	& 0.1 MJy sr$^{-1}$&  12 MJy sr$^{-1}$\\
Zodiacal Flux Density ($I_{LZ} \Omega_{tel}$)& 5 nJy& 220 $\mu$Jy\\
Stellar Magnitude$^a$  	& V=4.5 mag (60 Jy)&  [12]=3.0 mag (1.7 Jy)\\
Stellar Rejection		&10$^{-10}$	&10$^{-5}$\\
Stellar Leakage Signal 	& 6 nJy	& 17 $\mu$Jy\\
\hline
Planet Brightness 	& 6 nJy	& 0.3 $\mu$Jy \\
\enddata
\tablenotetext{a}{for a Solar twin at 10 pc.} 
\end{deluxetable}

\begin{deluxetable}{ccc}  
\tablecaption{Predicted Dust Emission at  10  $\mu$m \label{TPFexcesstable}}  
\tablehead{Star &  F$_\nu$ & Exo-Zodi$^a$ \\
 &   (mJy) & (Solar System = 1)} 
\startdata
HD 25998  & 0.041 &1.7\\
HD 40136  & 5.1   &42\\
HD 109085 & 23    &325\\
HD 199260 & 0.024 &1.1\\
HD 219482 & 0.004 &0.2\\
\enddata 
\tablenotetext{a}{$\ld$ at 10 $\mu$m in units of $10^{-7}$,
corresponding roughly to that of the Solar System \citep{backman93}. 
See eq.~[2] of \citet{Beichman06irs}.}
\end{deluxetable}

\clearpage
\pagestyle{plaintop}

\begin{figure}
\begin{center}
\includegraphics[width=4.7in,angle=-90]{f5.ps} 
\end{center}
\caption{Average \24um color relative to 2MASS $K_s$ band (2.16 \micron)
as a function of spectral type. Stars with excess emission or with
poor $K_s$ measurements are excluded. Error bars indicate the error on
the mean value within each bin (not the overall dispersion). Stellar
colors from the \citet{Bryden06a} F5-K5 survey and the
\citet{Gautier06} M star survey are also shown for comparison. The
trend is relatively flat over most of the range, with significantly
red colors only seen among the M type stars.} 
\label{kv24}
\end{figure}

\begin{figure}
\begin{center}
\end{center}
\caption{
Composite image of the field surrounding HD~105211
(marked by a plus sign). In addition to our MIPS \24um (green) and
\70um (red) images, the 2MASS $K_s$ band image is overlaid in blue.
While dim background stars show up as blue points, 
the cool Mira variable star CL Cru (triangle) has strong \24um emission
with a much broader PSF, resulting in an overall green color.
This bright star comtaminated the broad IRAS scanning beam (dashed
yellow rectangle) prohibiting the detection of dust around HD~105211.
In the MIPS image, HD~105211 is well resolved, showing strong excess
emission at \70um.
The neighoring star CPD-63 2145B (asterisk), while detected at \24um,
does not give off significant emission at \70um.} 
\label{MIPSimage}
\end{figure}

\begin{figure}
\begin{center}
\includegraphics[width=4.8in,angle=-90]{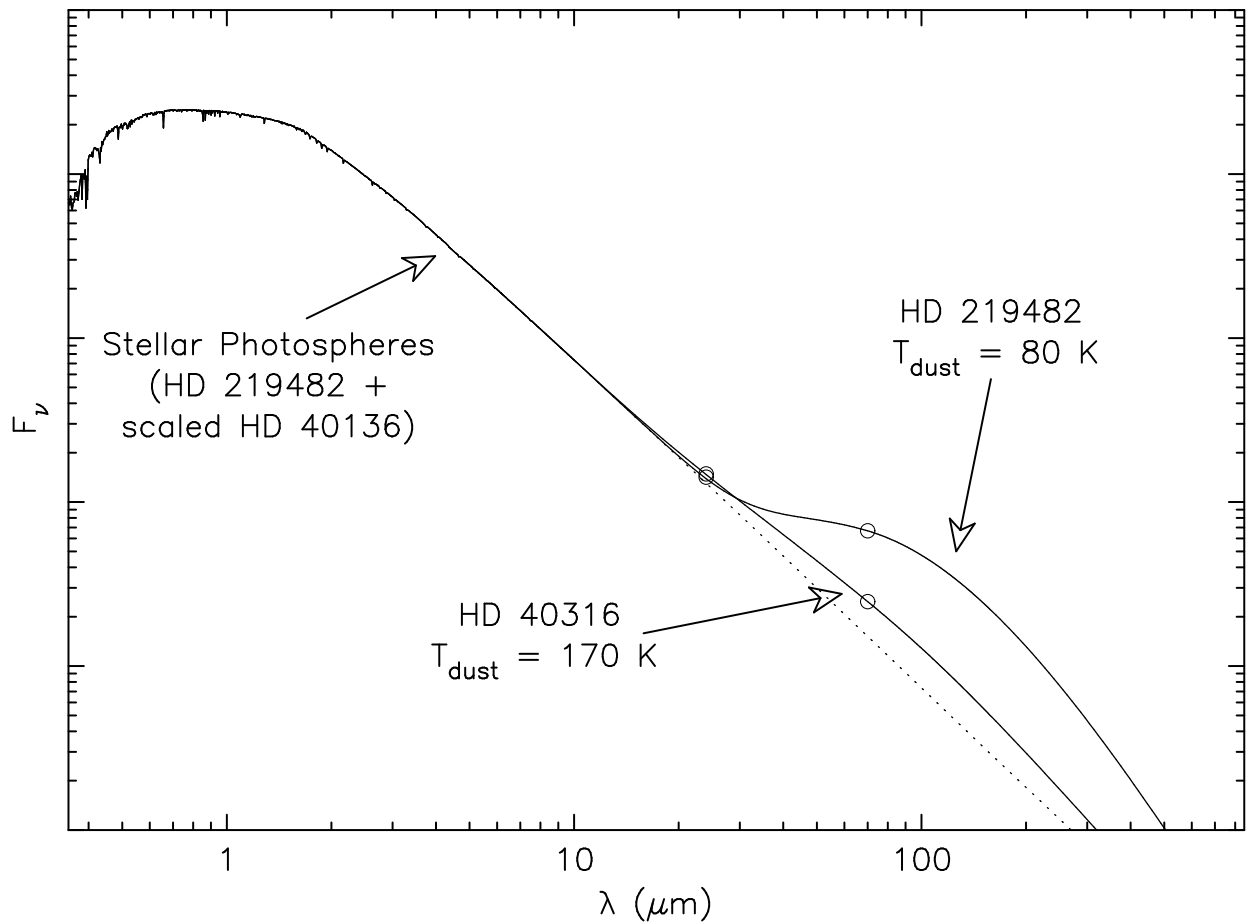} 
\end{center}
\caption{Spectral energy distributions for two stars with
IR excess at both 24 and \70um. 
The emission from the two stars has been scaled such that their
photospheres overlap, emphasizing the difference in far-IR emission.
The observed fluxes at 
each wavelength are shown as open circles that are fit with a
combination of emission from the stellar photosphere ({\it dotted line})
and from orbiting dust. The dust emission of HD 40316 is fit with 170 K dust, whereas HD 219482, with stronger \70um and weaker \24um emission, is fit with
cooler, 80 K dust.} 
\label{sed}
\end{figure}

\begin{figure}
\begin{center}
\includegraphics[width=4.7in,angle=-90]{f11.ps} 
\end{center}
\caption{Dust temperature for stars with IR excess in this sample and
from \citet{Bryden06a}. For stars with excess measured at both 24 and
\70um (solid points), the dust SED is fit with a representative
temperature. Those stars with a single measurement of excess at \70um
only have (3-$\sigma$) upper limits. 
With some assumptions for the grain properties, the dust temperatures
can be translated to orbital distances.
Several lines of constant distance are shown for comparison.
The observed systems with measured dust temperatures are mostly 
consistent with large blackbodies orbiting at $\sim$10 AU or
small, low-emissivity grains at $\sim$100 AU. 
There is no clear evidence for orbital distance changing as a function
of spectral type.} 
\label{Tbytype}
\end{figure}

\end{document}